\documentclass[11pt]{article}
\usepackage{jheppub}
\usepackage{epstopdf}
\usepackage{enumitem}
\usepackage{tcolorbox}

\usepackage{jheppub}
\usepackage{mathrsfs}
\usepackage{psfrag}
\usepackage{color}
\usepackage{hyperref}

\usepackage{slashed}
\usepackage{feynmp-auto}
\usepackage{simplewick}
\usepackage{cancel}

\usepackage{todonotes}

\usepackage{amsmath}
\usepackage{amsfonts}
\usepackage{graphicx}
\usepackage{amssymb}
\usepackage{xcolor}

\usepackage{mathtools}

\usepackage{amsmath,bbm,array,amsfonts,graphicx,wrapfig,arydshln,lscape,float,multirow,longtable,rotating,makecell}
\usepackage{url}

\DeclareMathAlphabet\mathbfcal{OMS}{cmsy}{b}{n}

\newcommand{\ra}{\rangle}

\newcommand{\p}{\partial}
\newcommand{\ve}{\varepsilon}
\newcommand{\la}{\langle}

\newcommand{\eq}{\begin{equation}}
\newcommand{\eqe}{\end{equation}}
\newcommand{\eqa}{\begin{eqnarray}}
\newcommand{\eqae}{\end{eqnarray}}
\newcommand{\nn}{\nonumber}
\newcommand{\bn}{\begin{enumerate}}
\newcommand{\en}{\end{enumerate}}

\newcommand{\eqc}[1]{(\ref{#1})}

\def\identity{{\rlap{1} \hskip 1.6pt \hbox{1}}}
\def\iden{\identity}

\def\CO{{\mathcal O}}

\def\IC{\mathbb{C}}

\def\CA{{\mathcal A}}

\def\CL{{\mathcal L}}
\def\CM{{\mathcal M}}

\def\CO{{\mathcal O}}

\def\a{\alpha}
\def\b{\beta}
\def\g{\gamma}
\def\e{\epsilon}
\def\ve{\varepsilon}
\def\z{\zeta}

\def\k{\kappa}
\def\l{\lambda}
\def\m{\mu}
\def\n{\nu}

\def\r{\rho}

\def\s{\sigma}

\def\G{\Gamma}
\def\D{\Delta}

\def\O{\Omega}

\def\half{\frac{1}{2}}

\def\p{\partial}

\def\identity{{\rlap{1} \hskip 1.6pt \hbox{1}}}

\newcommand{\bfig}{\begin{figure}}
\newcommand{\efig}{\end{figure}}

\def\abs#1{{\left| #1 \right|}}

\def\bl#1\el{\begin{align} #1 \end{align}}
\def\bg#1\eg{\begin{gather} #1 \end{gather}}

\def\bld#1\eld{\begin{aligned} #1 \end{aligned}}
\def\bgd#1\egd{\begin{gathered} #1 \end{gathered}}

\newcommand{\bra}[1]{\langle{#1}|}
\newcommand{\ket}[1]{|{#1}\rangle}

\newcommand{\sbra}[1]{ [{#1} |}
\newcommand{\sket}[1]{ | {#1} ]}

\renewcommand{\bf}{\textbf}

\newcommand{\AB}[1]{\langle #1 \rangle}
\newcommand{\SB}[1]{[ #1 ]}
\newcommand{\MixLeft}[3]{\langle #1 | #2 | #3 ]}
\newcommand{\BS}[1]{\boldsymbol{#1}}

\newcommand{\RN}[1]{%
  \textup{\uppercase\expandafter{\romannumeral#1}}%
}

\title{Classical potential for general spinning bodies}
\author{Ming-Zhi Chung$^{1}$}
\author{Yu-tin Huang$^{1,2}$}
\author{Jung-Wook Kim$^{3}$}

\affiliation{$^1$ Department of Physics and Astronomy, National Taiwan University, Taipei 10617, Taiwan}
\affiliation{$^2$ Physics Division, National Center for Theoretical Sciences, National Tsing-Hua University, No.101, Section 2, Kuang-Fu Road, Hsinchu, Taiwan}
\affiliation{$^3$ Department of Physics and Astronomy, Seoul National University, Seoul 08826, Korea
}

\emailAdd{dchung0741@gmail.com}
\emailAdd{yutinyt@gmail.com}
\emailAdd{jwkonline@snu.ac.kr}

\abstract{In this paper we compute the spin-dependent terms of the gravitational potential for general spinning bodies at the leading Newton's constant $G$ and to all orders in spin. We utilize the on-shell approach, which extracts the classical potential directly from the scattering amplitude. For spinning particles, extra care is required due to the fact that the spin space of each particle is independent. Once the appropriate matching procedures are applied, taking the classical-spin limit we obtain the potential for general spinning bodies. When the Wilson coefficients are set to unity, we successfully reproduced the potential for the Kerr black hole. Interestingly, for finite spins, we find that the finite-spin deviations from Kerr Wilson coefficients cancel with that in the matching procedure, reproducing the Kerr potential without the need for taking the classical-spin limit. Finally, we find that when cast into the chiral basis, the spin-dependence of minimal coupling exhibits factorization, allowing us to take the classical-spin limit straight forwardly.       }

\begin{document}
\begin{flushright}
\vspace{10pt} \hfill{NCTS-TH/1907} \vspace{20mm}
\end{flushright}
\maketitle

\section{Introduction}
There has been a long history of development in the extraction of classical quantities from observables of quantum field theory. Earlier examples include perturbative computations of the metric from vacuum expectation value of the gravitational field~\cite{Duff:1973zz}, or the stress-tensor form factors~\cite{Donoghue:2001qc, BjerrumBohr:2002ks}. Later on, it was shown that two-body potentials can also be extracted from scattering amplitudes~\cite{Neill:2013wsa, Cheung:2018wkq, Bjerrum-Bohr:2018xdl}. Recently there has been a surge of renewed interest inspired by the successful extension of the modern advancements of scattering amplitudes to these problems. For example, the simplification of loop-level gravitational scattering amplitudes either through the double copy~\cite{Kawai:1985xq} or BCJ relations~\cite{Bern:2008qj}, has been utilized for the computation of classical potentials~\cite{Bern:2019nnu, Bern:2019crd}. Furthermore, the massive spinor-helicity formalism introduced by one of the authors~\cite{Arkani-Hamed:2017jhn}, has enabled a more streamlined approach to the computation of spin-effects in the classical potential~\cite{Guevara:2017csg, Chung:2018kqs} and scattering angle~\cite{Guevara:2018wpp}.

An important aspect in the success of applying amplitudes for black hole physics is that for long range forces, a black hole is well described as a point particle. In essence, this is a reflection of the no hair theorem: for the asymptotic observer, a black hole is completely characterized by its mass, spin and charge, much like that of an elementary particle. Indeed, this feature is well appreciated in the context of Schwarzschild black holes, being treated as minimally (gravitationally) coupled massive scalar. For spinning black holes, the natural counterpart would be spinning particles. Indeed, the new massive spinor-helicity formalism introduced in~\cite{Arkani-Hamed:2017jhn}, allows one to kinematically identify the minimally coupled spin-$s$ particle. In particular, its cubic coupling to the graviton yields the following three-point amplitude:
\eq\label{Intro1}
x^2\frac{\kappa}{2}\frac{\langle \mathbf{12}\rangle^{2s}}{m^{2s{-}1}}
\eqe
where the angle brackets represents the spinors of the massive legs and the definition of $x$ is of kinematic origin:
\eq\label{xDef}
x\lambda_{q}^{\alpha}=\frac{\tilde{\lambda}_{q \dot\alpha} p_{1}^{\dot{\alpha}{\alpha}}}{m}\,,
\eqe 
where $\lambda_{q}, \tilde{\lambda}_{q}$ are the spinors of the massless leg. For $s\leq\frac{3}{2}$ this amplitude matches to that of minimal coupling. Remarkably, Kerr black hole can  be identified as the classical-spin limit of such minimal coupling, i.e. $s\rightarrow \infty, \hbar\rightarrow 0$ with $s\hbar$ fixed. Indeed this was confirmed in the matching of the three-point amplitude induced from minimal coupling with the worldline formalism with black hole Wilson coefficients~\cite{Chung:2018kqs} or the coupling to the Kerr black hole  stress-tensor~\cite{Guevara:2018wpp}, as well as through its effect on the impulse~\cite{Arkani-Hamed:2019ymq}.

The simplicity of on-shell approach for Kerr black holes motivates us to apply it to the spin effects of general stellar objects.  A general purpose approach is well established in the context of the one-particle effective field theory (EFT), where one works with a worldline action in a non-trivial background~\cite{Goldberger:2004jt,Porto:2005ac,Porto:2006bt,Porto:2008tb,Porto:2008jj,Levi:2008nh}. The interacting part of the action can be organized as 
\eq
\mathcal{L}_{Int}=\sum_{a}\,C_{a} \mathcal{O}_a\,,
\eqe
where $\mathcal{O}_a$ is comprised of the worldline fields as well as curvature tensor of the background sourced by the worldline, and $C_{a}$s are the ``Wilson coefficients". Distinct objects are then reflected in their distinct value for these Wilson coefficients. The potential is then derived by treating the operators as sources, exchanging quanta of gravitational fields. 

In this paper, we derive the leading spin effect terms of the classical potential for general spinning objects by constructing the amplitude associated with the general one-particle EFT. The closed form of the EFT for spin couplings to all order in spin can be found in~\cite{Levi:2015msa}. We begin by first converting these worldline operators, linear in gravitational field strength, to a three-point amplitude. The worldline operators act on the physical Hilbert space, whose states are momentum eigenstates and form irreducible representations of the massive little group, $SU(2)$. Thus each operator is understood as a matrix acting in little group space, i.e. $\mathcal{O}^{\{I\}}\,_{\{J\}}$ where $\{I\}, \{J\}$ are irreps of some particle, chosen to be 1 without loss of generality. A prescription for obtaining an amplitude from a set of operators in little group space, is to sandwich it with polarization tensors of particle 1 and 2; schematically,
\eq\label{EFTAmp}
\sum_a\,C_{a}\mathcal{O}_a^{\{I\}}\,_{\{J\}}\;\;\rightarrow \;\;\sum_a\,C_{a}(\varepsilon^{*\{\mu_s\}}_2)^{\{I\}}\mathcal{O}_a^{\{K\}}\,_{\{J\}}(\varepsilon_{1,\{\mu_s\}})_{\{K\}}= M_3(1_s^{\{J\}}2_s^{\{I\}}q)\quad \vcenter{\hbox{\includegraphics[scale=0.5]{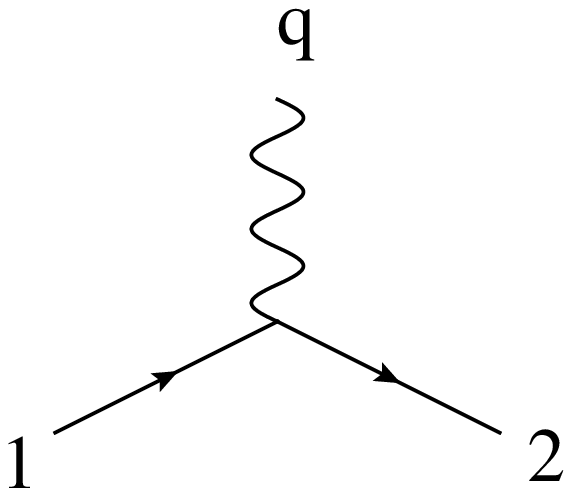}}}\,
\eqe
where $\{\mu_s\}$ is the Lorentz indices of the polarization tensor and $\{ I \}$, $\{J\}$ are little group indices for the polarization tensors of two distinct particles. Importantly, the polarization tensors themselves carry non-trivial spin-effects. Indeed setting up the polarization tensor $\varepsilon_{\{\nu_s\}}^{\{I\}}(p_0)$ for reference momentum $p_0$, all other polarization tensors can be obtained from it via: 
\eq
\varepsilon_{\{\mu_s\}}^{\{I\}}(p_i)=\left[G(p_i,p_0)\right]_{\{\mu_s\}}\,^{\{\nu_s\}}\varepsilon_{\{\nu_s\}}^{\{I\}}(p_0)
\eqe
where $G(p_i,p_0)$ is the Lorentz boost that transform $p_0$ to $p_i$. The leading order in $G$ (1 PM) gravitational potential is then encoded in the factorization limit of the four-point amplitude, whose residue is the product of the aforementioned three-point amplitude.

Importantly, in the extraction of the classical potential, one must take into account the spin degrees of freedom that are inherent in the polarization tensors. To this end, we map the out-going polarization tensor to the incoming one, 
\eq
\varepsilon_{\{\mu_s\}}^{*\{I\}}(p_{out})=\varepsilon_{\{\nu_s\}}^{*\{I\}}(p_{in})\left[G(p_{in},p_0)G^{-1}(p_{out},p_0)\right]^{\{\nu_s\}}\,_{\{\mu_s\}}\,,
\eqe 
and the spin-dependent pieces are contained in $G(p_{in},p_0)G^{-1}(p_{out},p_0)$. We introduce \emph{Hilbert space matching} as a procedure for incorporating the effects of such a succession of Lorentz rotations, which can be decomposed into pure boosts and pure rotations. The rotation part, which is explored in section \ref{sec:HilbertMatch}, depends on the reference momentum $p_0$, whose choice is observed to be related to \textit{spin supplementary conditions} (SSC). For our two-body problem, a natural choice for $p_0$ is the C.O.M momenta, whose results also matches with the Newton-Wigner SSC where the spin operators satisfies the canonical commutation relations~\cite{Levi:2015msa}. The boost part, which is explored in section \ref{sec:MinCoupBH}, is independent of reference momentum $p_0$ and affects the interpretation of Wilson coefficients with vanishing effects in the classical-spin limit.

To simplify the discussions, we first consider the classical-spin limit and construct the leading post-Newtonian (PN) order potential to all orders in spin-operators for general spinning compact bodies, which is the leading term in the double expansion of $G$ and $p^2$. The result is first checked by comparing with known quartic order in spin results for general spinning compact bodies~\cite{tulczyjew1959equations,Barker:1975ae,Hergt:2008jn,Levi:2010zu,Levi:2014gsa}, and then by comparing with equivalent order potential for binary black holes by Vines and Steinhoff~\cite{Vines:2016qwa}.

Next, we analyse the finite-spin effects. Here by finite-spins we are referring to keeping $s$ fixed, absorbing a factor of $\hbar$ into $s$ while setting the remaining $\hbar$s to zero for the classical-limit. Note that the operator that enters the final potential is in effect given by,
\eq
\left(\mathcal{O}_{eff}\right)^{\{I\}}\,_{\{J\}}\equiv (\varepsilon^{*}_{1,\{\nu_s\}})^{\{I\}} \left[ G(p_{in},p_0)G^{-1}(p_{out},p_0) \right]^{\{\n_s\}}_{~~\{\m_s\}} \mathcal{O}^{\{K\}}\,_{\{J\}} (\varepsilon_{1}^{\{\mu_s\}})_{\{K\}} \,.
\eqe
It is interesting to consider the Wilson coefficients in such \textit{effective operator basis}. Remarkably, we find that when minimal coupling is recast into this effective basis, the ``effective" Wilson coefficients are simply 1 without taking the classcial-spin limit!\footnote{An equivalent conclusion has been reached independently from heavy particle effective theory (HPET) point of view~\cite{Aoude:2020onz} while this manuscript was under revision.} Recall that in~\cite{Chung:2018kqs}, reading off the Wilson coefficients associated with minimal coupling from eq.(\ref{EFTAmp}) yields
\eq
C_{S^n}=1+\mathcal{O}(s^{-1})\,,
\eqe
i.e. there are deviations from the  Kerr BH that vanishes in the classical-spin limit. The fact that the effective Wilson coefficient is $1$ indicates that the ``finite-spin" effects are exactly cancelled by the finite spin terms in the Hilbert-space matching procedure.\footnote{The same statement has been made in the work of~\cite{Vaidya:2014kza}. However, the spin-operator defined there are different than that in this work. Furthermore, the result disagrees with the earlier work of~\cite{Holstein:2008sx}. We will comment on these discrepancies in detail. }

Finally, we study in detail the classical-spin limit of minimal coupling. Focusing on the spin-dependence of the amplitude, we find that

\noindent \begin{center} \framebox{ \parbox{0.9\textwidth}{{\textit{when cast in the chiral basis, the one graviton exchange for minimal coupling factorizes completely into a spin-dependent combinatoric factor and a spin-independent kinematic term.}}} } \end{center}

\noindent This ``\textit{universality}", allows us to obtain the classical-spin limit of minimal coupling from any finite-spin computation, by simply retaining the universal piece and replacing the combinatoric factors by their infinite-spin asymptotic form. This turns out to be the prescription presented in~\cite{Chung:2018kqs}. 
We also show that universality is a reflection of binomial expansion hidden in the amplitude, and we comment on its persistence at one-loop order in appendix~\ref{sec:2PMUniv} and its generalisation to non-minimal couplings in appendix~\ref{sec:resrep}.

This paper is organized as follows. In section \ref{sec:EFT3pt}, we review the matching of one-particle EFT to on-shell three-point amplitudes. Next, we compute the leading PN classical potential between two bodies to all orders in spin in section \ref{sec:1PMPot}. After presenting 
Hilbert space matching in section \ref{sec:FinSpin}, the justification for the prescription given in \cite{Chung:2018kqs} is outlined in section \ref{sec:1PMrevisit} for tree-level(1 PM) order. 
We conclude our paper with section \ref{sec:conc}.

\newpage
 
\section{One-particle effective action to on-shell amplitudes} \label{sec:EFT3pt}
In this section, we derive a map between the one-particle EFT and the three-point amplitude. Since residue of the one-graviton exchange in the four-point amplitude is given by the product of two three-point amplitudes, the latter contains all necessary information to compute the leading order potential.

\subsection{Three-point amplitude of general EFT}
We begin by considering the effective action of a classical point particle coupled to a quantum gravitational background. 
Such a formulation has been introduced by Goldberger and Rothstein~\cite{Goldberger:2004jt} to compute relativistic corrections to Newtonian potential, and the first attempt to include spin and multipole moments has appeared by Porto~\cite{Porto:2005ac}. The formulation we base our construction on was introduced by Levi and Steinhoff~\cite{Levi:2015msa}, where a generic treatment of rotational variables were introduced and a closed form description of spin couplings to all orders in spin were given. Currently the approach has become one of the main techniques for computing the spin-dependent post-Newtonian effects of gravity~\cite{Porto:2005ac,Porto:2006bt,Porto:2008tb,Porto:2008jj,Levi:2008nh,Perrodin:2010dy, Porto:2010tr,Levi:2010zu,Levi:2014gsa,Levi:2015msa,Levi:2017kzq,Levi:2019kgk}; consult the reviews~\cite{Porto:2016pyg,Levi:2018nxp} for a more complete list of references. This is an effective action where the gravitational field is decomposed into modes with different scaling properties and modes shorter than the scale $r_s$ of the compact object has been removed, thus allowing us to approximate the black hole as an isolated compact object. One then starts with the following worldline action~\cite{Porto:2005ac}:
\eq\label{1bdy}
S=\int d\sigma \;\left\{-m\sqrt{u^2}-\frac{1}{2}S_{\mu\nu}\Omega^{\mu\nu}+L_{SI}\left[u^\mu,S_{\mu\nu}, g_{\mu\nu}(y^\mu)\right]\right\}
\eqe
where $u^\mu\equiv \frac{dy^\mu}{d\sigma}$, $S_{\mu\nu}$ correspond to the spin-operator, and $\Omega_{\mu\nu}$ is the angular velocity. In this section we choose the covariant Spin Supplementary Condition (SSC) $p^\mu S_{\mu\nu}=0$, where $p^\mu=\frac{(p_1-p_2)_\nu}{2}$.\footnote{The difference $\frac{p_1 - p_2}{2}$ was used to define average momentum to comply with conventions of amplitude literature; all momenta are considered to be incoming.} This allows us to identify the spin-operator with the spin-vector via $S^{\mu\nu}=-\frac{1}{m}\epsilon^{\mu\nu\rho\sigma}p_\rho S_\sigma$. Note however, that the choice of SSC does not affect the on-shell three-point amplitude as we will show shortly in section~\ref{sec:SSC}.

The first two terms of the EFT Lagrangian eq.\eqc{1bdy} are called minimal coupling and are universal, irrespective of the details of the 
point-like particle, while the terms in $L_{SI}$ correspond to spin-induced multipole terms that are beyond minimal coupling, and depend on the 
inner structure of the particle. The angular velocity $\O^{\m\n}$ is defined as $\O^{\m\n} := e^\m_A \frac{D e^{A\n}}{D\s}$, where $e^{\m}_A (\s)$ is the tetrad attached to the worldline of the particle. The spin-induced multipole moments given in~\cite{Porto:2005ac, Porto:2008tb, Porto:2008jj, Levi:2015msa} is:
\bl
\bld
L_{SI} &= \sum_{n=1}^{\infty} \frac{(-1)^n}{(2n)!} \frac{C_{\text{ES}^{2n}} }{m^{2n-1}} D_{\m_{2n}} \cdots D_{\m_{3}} \frac{E_{\m_{1} \m_{2}} }{\sqrt{u^2}} S^{\m_{1}} S^{\m_{2}} \cdots S^{\m_{2n-1}} S^{\m_{2n}}
\\ &\phantom{=} + \sum_{n=1}^{\infty} \frac{(-1)^n}{(2n+1)!} \frac{C_{\text{BS}^{2n+1}} }{m^{2n}} D_{\m_{2n+1}} \cdots D_{\m_{3}} \frac{B_{\m_{1} \m_{2}} }{\sqrt{u^2}} S^{\m_{1}} S^{\m_{2}} \cdots S^{\m_{2n}} S^{\m_{2n+1}}\,.
\eld \label{eq:obEFTLag}
\el
where $E$ and $B$ are the electric and magnetic components of the Weyl tensor\footnote{The vacuum Einstein equation reduces to $R_{\m\n} = 0$, therefore the Riemann tensor is equal to the Weyl tensor $R_{\m\n\l\s} = C_{\m\n\l\s}$ in this background.} defined as:
\bl
E_{\m\n} &:= R_{\m\a\n\b} u^{\a} u^{\b}
\nonumber\\ B_{\m\n} &:= \half \e_{\a\b\g\m} R^{\a\b}_{~~\delta \n} u^{\g} u^{\delta}\,,
\el
and the covariant derivatives act on the Riemann tensors. Here the Riemann tensors contain linear perturbations around flat space, and the information with regards to non-trivial backgrounds is encoded in the Wilson coefficients $C_{\text{S}^n}$, which is set to $1$ for Kerr black-holes.

Since each Riemann tensor is linear in the perturbed metric, these operators represent the coupling of the worldline particle to a graviton. Thus it naturally maps to a three-point amplitude involving two identical massive spin-$s$ state and the emission of a graviton:
\eq
\vcenter{\hbox{\includegraphics[scale=0.5]{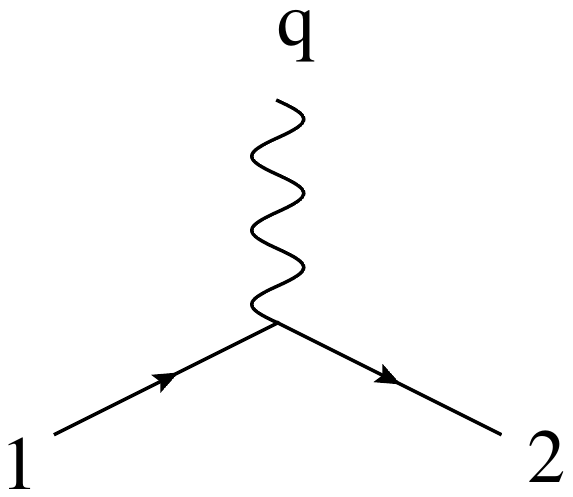}}}\quad=M^{-2}_s=\sum_{i}\; \varepsilon^{\ast \{\mu_s\}}_{2,\{I_s\}}\, \mathcal{O}_i\, \varepsilon_{1,\{\mu_s\}}^{\{J_s\}}\,, \label{eq:3ptGenForm}
\eqe
where $\varepsilon_{1}^{\{\mu_s\},\{J_s\}}, \varepsilon^{\{\mu_s\},\{I_s\}}_{2}$ are the polarization tensors of particles $1,2$ respectively, with $\{\mu_s\}$ representing the symmetrized $s$ Lorentz indices, and $\{I_s\}$ represent the $2s$ symmetrized $SU(2)$ little group indices\footnote{For brevity, equations appearing in section~\ref{sec:1PMPot} and onwards will simply denote little group indices as uppercase Latin indices without curly brackets, e.g. $I$, $J$, and $K$.}. These polarisation tensors act as basis vectors in the little group space. The task is then simply working out the action of the worldline operators on the physical states. Note that as the worldline operators $\mathcal{O}_i$ are Lorentz scalars, the Lorentz indices of the polarization tensors of particle 1 and 2 are contracted with each other. The worldline derivatives are converted as:
\eq
u^\mu=\frac{p_1^\mu}{m},\;\; D^\mu\rightarrow \partial^\mu=-iq^\mu,\;\; h_{\mu\nu}=\varepsilon_{q,\mu}\varepsilon_{q,\nu}\,.
\eqe

Having converted the spin-operator to spin-vectors through covariant SSC, naively one then simply identify the latter with the Pauli-Lubanski pseudo-vector,  
\eq\label{PauliDef}
 S^\mu = - \frac{1}{2m}\epsilon^{\mu\nu\rho\sigma} p_\nu J_{\rho\sigma}\,,
\eqe
where $P$ acts on the physical state which we choose to be particle $1$, hence $p^\mu=p^\mu_1$.\footnote{Such a spin-vector operator with operator $P$ substituted by an eigenvalue $p_1$ will sometimes be denoted as $S^\m (p_1)$, although this dependence will be usually implicit in our notations.} However, as physical states are irreps of the $SU(2)$ little group, the spin operators in $\mathcal{O}_i$ should be thought of as operators acting on the little group space. To relate $S^\m$ defined in eq.(\ref{PauliDef}) into a little group space operator, we sandwich it with polarization tensor of say particle $1$. That is, we define:  
\bg
(\mathbb{S}^\mu)_{\{I_s\}}\,^{\{J_s\}}\equiv\varepsilon_{1,\{I_s\}}^\ast \,S^\mu\, \varepsilon_1^{\{J_s\}} \Leftrightarrow S^\mu\, \varepsilon_1^{\{J_s\}} \equiv \sum_{\{I_s\}} (\mathbb{S}^\mu)_{\{I_s\}}\,^{\{J_s\}} \varepsilon_1^{\{I_s\}} \label{eq:OpAction}
\eg
where we've suppress the Lorentz indices of the two $\varepsilon_{1,\{I_s\}}$, which are contracted with Lorentz indices of $J$ in eq.(\ref{PauliDef}), and  $\varepsilon_{1,\{\m_s\},\{I_s\}}^\ast \varepsilon_1^{\{\m_s\},\{J_s\}} = \delta^{\{J_s\}}_{\{I_s\}}$. This definition is equivalent to the definition considered in~\cite{Maybee:2019jus} since spinor-helicity variables form a representation of the Lorentz group, therefore the variables can be considered as one-particle states.
\bl
\varepsilon_{1,\{I_s\}}^\ast \,S^\mu\, \varepsilon_1^{\{J_s\}} \Leftrightarrow \bra{p_1, \{I_s\} } S^\mu \ket{p_1, \{J_s\} }
\el
This means it is unnecessary to compute Noether currents as in~\cite{Maybee:2019jus} to obtain matrix elements of spin variables. Details of this discussion will be given in section \ref{sec:SysPolSpin}. As such, $\mathbb{S}^\mu$ is a Lorentz vector carrying $2s\otimes 2s$ $SU(2)$ indices of particle 1, separately symmetrized.

Now, the spin-vectors $\mathbb{S}^\mu$ appear in the one-particle EFT as
\eq
\cdots (q\cdot \mathbb{S})_{\{I_s\}}\,^{\{J_s\}} (q\cdot \mathbb{S})_{\{J_s\}}\,^{\{K_s\}}\cdots=\cdots\varepsilon_{1\{I_s\}}^\ast \,(q\cdot S)\, \varepsilon_1^{\{J_s\}}\varepsilon_{1\{J_s\}}^\ast \,(q\cdot S)\, \varepsilon_1^{\{K_s\}}\cdots\,,
\eqe
where the equality corresponds to substituting eq.(\ref{eq:OpAction}). We see that it is equivalent to products of $SL(2,\IC)$ operators, sandwiched with the polarization tensors contracted in their little-group indices. Each contraction yields a projection operator
\eq\label{ProjectP}
\varepsilon_{\alpha_1\dot{\alpha}_1\cdots \alpha_s\dot{\alpha}_s}^{\{J_s\}}\varepsilon_{\{J_s\},\beta_1\dot{\beta}_1\cdots \beta_s\dot{\beta}_s}^\ast=\mathbb{P}_{\alpha_1\beta_1\cdots \alpha_s\beta_s;\dot\alpha_1\dot\beta_1\cdots \dot\alpha_s\dot\beta_s}=\left(\mathbb{I}-\mathbb{\hat{P}}\right)_{\alpha_1\beta_1\cdots \alpha_s\beta_s;\dot\alpha_1\dot\beta_1\cdots \dot\alpha_s\dot\beta_s}
\eqe
where $\mathbb{P}$ is identified with a transverse projection operator, with $p_1^{\alpha_1\dot{\alpha}_1}\mathbb{P}_{\alpha_1\beta_1\cdots \alpha_s\beta_s;\dot\alpha_1\dot\beta_1\cdots \dot\alpha_s\dot\beta_s}=0$, and $\mathbb{\hat{P}}$ is comprised of products of $p_1/m$ and the Levi-Cevita tensors $\epsilon$, ensuring the transverse property of $\mathbb{P}$.  As we will now show, due to the special three-point kinematics, the $\mathbb{\hat{P}}$ part of the projection operator will not contribute to the final amplitude.

First, we derive the explicit form of $(q\cdot S)$.  In $SL(2,\IC)$ representation, the Lorentz generator $J_{\mu\nu}$ splits into chiral and anti-chiral representations. It's action on an $(s,0)$ representation can be written as:
\eq
(J_{\mu\nu})_{\alpha_1\alpha_2\cdots\alpha_{2s}}\,^{\beta_1\beta_2\cdots\beta_{2s}}=\sum_{i}(J_{\mu\nu})_{\alpha_i}\,^{\beta_i}\,\bar{\mathbb{I}}_i \stackrel{\cdot}{=} 2s(J_{\mu\nu})_{\alpha_1}\,^{\beta_1}\,\bar{\mathbb{I}}_1,\quad (J_{\mu\nu})_{\alpha}\,^{\beta}=\frac{i}{2}\left(\sigma_{[\mu}\bar{\sigma}_{\nu]}\right)_{\a}^{~\b}\,, \label{eq:LorGenDef}
\eqe
where $\bar{\mathbb{I}}_i=\delta_{\alpha_1}^{\beta_1}\cdots \delta_{\alpha_{i{-}1}}^{\beta_{i{-}1}}\delta_{\alpha_{i{+}1}}^{\beta_{i{+}1}}\cdots \delta_{\alpha_{2s}}^{\beta_{2s}}$, with a similar form for the conjugate representation. The sign $\stackrel{\cdot}{=}$ means the RHS can be used instead of LHS of $\stackrel{\cdot}{=}$ as $SL(2,\IC)$ indices are symmetrised, but the proper definition for $J_{\m\n}$ is the expression between $=$ and $\stackrel{\cdot}{=}$. Using this, we find that
\bl
m \left( S_\m \right)_{\a}^{~\b} &= \frac{1}{4} \left[ \s_\m (p_1 \cdot \bar{\s}) - (p_1 \cdot \s) \bar{\s}_\m \right]_{\a}^{~\b} \,, \label{eq:spinopangsqdefs}
\\ m \left( S_\m \right)^{\dot{\a}}_{~\dot{\b}} &= - \frac{1}{4} \left[ \bar{\s}_\m (p_1 \cdot \s) - (p_1 \cdot \bar{\s}) \s_\m \right]^{\dot{\a}}_{~\dot{\b}} \,.
\label{eq:spinopsq}
\el
When contracted with the massless momentum $q$, one finds: 
\bl
\bld
(q\cdot S)_{\a}^{~\b} &= \frac{x}{2} \lambda_{q\a} \lambda_q^\b\equiv \frac{x}{2} |q\rangle\langle q| 
\\ (q\cdot S)^{\dot\a}_{~\dot\b} &=-\frac{\tilde{\lambda}_{q}^{\dot\a} \tilde{\lambda}_{q\dot\b}}{2x}\equiv - \frac{|q][q|}{2x}\,,
\eld \label{PSDef}
\el
where the variable $x$ is defined in eq.(\ref{xDef}). Now since for our on-shell three-point kinematics, 
\eq
[q|p_1|q\rangle=0,
\eqe
any factors of $p_1$ in between $(q\cdot S)$ must vanish. This implies that the factors of $\mathbb{\hat{P}}$ in the projection operator will drop out, thus leading to,  
\eqa
 &&(q\cdot \mathbb{S})_{\{I_s\}}\,^{\{J_s\}}(q\cdot \mathbb{S})_{\{J_s\}}\,^{\{K_s\}} \cdots (q\cdot \mathbb{S})_{\{K_s\}}\,^{\{L_s\}}= \varepsilon_{1,\{I_s\}}^\ast (q\cdot S)\mathbb{P}(q\cdot S)\mathbb{P}\cdots \mathbb{P}(q\cdot S)\varepsilon_{1}^{\{L_s\}}\nonumber\\
&&\quad\quad\quad\quad\quad\quad\quad\quad\quad\quad\quad\quad\quad\quad\quad\quad\quad\quad= \varepsilon_{1,\{I_s\}}^\ast (q\cdot S)(q\cdot S)\cdots(q\cdot S)\varepsilon_{1}^{\{L_s\}}
\eqae
where the product of $(q \cdot S)$ factors in the last line denotes contraction over $SL(2,\IC)$ indices. Returning to the expression for the amplitude eq.\eqc{eq:3ptGenForm}, we have:
\eqa
\ve^{\ast \{\mu_s\}}_{2,\{I_s\}} (\CO_i \ve_{1,\{\mu_s\}})^{\{J_s\}} &=& \ve_{2,\{I_s\}}^{\ast \{\mu_s\}}\left[  \left( \CO_i \right)_{\{K_s\}}\,^{\{J_s\}}\ve_{1\{\mu_s\}}^{\{K_s\}} \right]
\nonumber\\ &=& \ve_{2,\{I_s\}}^{\ast \{\mu_s\}} \left[ \ve_{1\{K_s\}}^\ast (q\cdot S)(q\cdot S)\cdots(q\cdot S)\ve_{1}^{\{J_s\}}\ve_{1\{\mu_s\}}^{\{K_s\}} \right]\nonumber\\
&=& \ve_{2, \{I_s\}}^{\ast} \mathbb{P}(q\cdot S)\cdots(q\cdot S) \ve_{1}^{\{J_s\}}
\eqae
where we've used 
eq.(\ref{ProjectP})
. Finally since due to three-point kinematics,
\eq
\ve_{2, \{I_s\}}^{\ast} p_1(q\cdot S)=-\ve_{2, \{I_s\}}^{\ast} p_2(q\cdot S)=0,
\eqe
when sandwiched between $\ve_{2,\{I_s\}}^\ast$ and $(q\cdot S)$, $\mathbb{P}=\mathbb{I}$, we see that when converted to scattering amplitudes, we can simply replace $(q\cdot \mathbb{S})$ in $\mathcal{O}$ by $(q\cdot S)$.

Putting everything together, we finally arrive at the three-point amplitude derived from the one-particle EFT~\cite{Chung:2018kqs}
\bl
M_s^{2 \eta} =  \ve_2^{\ast \{I_s\}} \left[ \sum_{n=0}^{2s} \frac{\k m x^{2\eta} }{2} \frac{C_{\text{S}^n}}{n!} \left( - \eta \frac{q \cdot S}{m} \right)^n \right] \ve_{1\{J_s\}}  \label{eq:1bd3ptAmp}
\el
for integer spin $s$, where $\eta = +1$ for positive helicity graviton and $\eta = -1$ for negative helicity graviton. Since this expression is a contraction between polarisation tensors of different momenta, the expression \emph{cannot} be interpreted as a matrix element in terms of spin operators, a point that we will come back to in sections \ref{sec:HilbertMatch} and \ref{sec:MinCoupBH}. Note that for fixed $s$, the polarization tensor is in $(\frac{s}{2},\frac{s}{2})$ representation containing $s$ chiral and $s$ anti-chiral $SL(2,\IC)$ indices, and thus can only transform non-trivially under at most $2s$ spin vector operators $S^\mu$. Thus $M^{+2}_s$  will receive contributions from the terms in the one-particle effective action with $S^n$ where $n\leq 2s$.   The first two Wilson coefficients are fixed as unity from the universal terms in eq.\eqc{1bdy}, while $C_{\text{S}^{n>1}}$ are simply the electric and magnetic Wilson coefficients $C_{\text{ES}^{2n}}$ and $C_{\text{BS}^{2n+1}}$ of eq.\eqc{eq:obEFTLag}. As the polarization tensors can be written in terms of products of massive spinors:
\eq
\epsilon^{\mu_1\mu_2\cdots \mu_s}\;\rightarrow\;  \epsilon^{\alpha_1\alpha_2\cdots \alpha_s\dot\alpha_1\dot\alpha_2\cdots \dot\alpha_s}=\frac{1}{m^s}\lambda^{\alpha_1\{I_1}\lambda^{\alpha_2I_2}\cdots \lambda^{\alpha_sI_s}\tilde\lambda^{\dot\alpha_1I_{s{+}1}}\tilde\lambda^{\dot\alpha_2I_{s{+}2}}\cdots \tilde\lambda^{\dot\alpha_sI_{2s}\}}\,,
\eqe
the three-point amplitude can be written purely in terms of kinematic variables as:
\eqa\label{EFT3pt}
&&M_s^{2
} =\sum_{a+b\leq s}\; \frac{\k m x^{2
} }{2} C_{\text{S}^{a+b}}  n^{s}_{a,b} \la \bold{2} \bold{1} \ra^{s-a} \left( - 
\frac{x \la \bold{2}q \ra \la q\bold{1} \ra}{2m} \right)^a [ \bold{2} \bold{1} ]^{s-b} \left( 
\frac{[ \bold{2}q ] [ q\bold{1} ]}{2mx} \right)^b,\nonumber\\
 &&\quad\quad\quad n^{s}_{a,b}\equiv\frac{1}{m^{2s}} {s \choose a} {s \choose b}
 \,.
\eqae
The coupling constant $\k$ is defined as $\sqrt{32 \pi G}$ where $G$ is the gravitational constant. Note that as the polarization tensors carry both dotted and un-dotted indices, the amplitude depends on both chiral and anti-chiral spinors. We will refer to such representation as the polarization basis. Using Dirac equations we can convert to a representation that is purely in terms of chiral spinors, which we will refer to as the chiral basis.

\subsection{The equivalence of Covariant and NW SSC at three-points} \label{sec:SSC}
We've derived the three-point amplitude from the one-particle EFT using the covariant SSC, obtaining eq.(\ref{EFT3pt}). At this point, one might worry that this implies that the three-point amplitude might be scheme dependent. This is not the case as we now show.

Switching SSC is equivalent to shifting the ``centre'' of the body~\cite{Levi:2015msa}. Let us start with covariant SSC and consider the difference in switching to NW SSC, defined as $S^{\m\n}(P_\n + m e_\n) = 0$. Consider the following shift from covariant SSC to a new SSC
\bg
S_{\text{cov}}^{\m\n} \to S^{\m\n} = S_{\text{cov}}^{\m\n} - X^\m P^\n + P^\m X^\n\,. \label{eq:SSCshift}
\eg
The following choice of $X^\m$ shifts from covariant SSC to NW SSC , where $e^\m$ is a unit time-like vector and $P^2 = m^2$.
\bl
X^\m &= \frac{S_{\text{cov}}^{\m\n} e_\n}{m + P \cdot e} \label{eq:Xvecdef}\,,
\el
The change in spin length $\half S^{\m\n} S_{\m\n}$ under this switching of SSC is $\half S^{\m\n} S_{\m\n} = \half S_{\text{cov}}^{\m\n} S_{\text{cov}\m\n} + m^2 X^2$. It may not be obvious that spatial displacement of the centre $X^2 < 0$ reduces spin length, but this is because the increase in mass dipole moment $S^{0i}$ is far greater than the increase in spin $S^{ij}$.

The changes in the three-point amplitude induced from switching SSC by eq.\eqc{eq:SSCshift} is proportional to the following expression.
\bl
\delta (\O_{\m\n} S^{\m\n}) \propto ( q_\m \e^{\pm}_\n - \e^{\pm}_\m q_\n ) X^\m P^\n\,,
\el
where $\e$ is the polarization vector whose square gives the polarization tensor of the graviton. Due to three-point kinematics $q \cdot P = 0$, only the first term needs to be considered. Thus we have 
\bl
\delta (\O_{\m\n} S^{\m\n}) \propto q_\m X^\m \propto S_{\text{cov}}^{\m\n} q_\m q_\n = 0 \,.
\el
where the last equality is obtained from substituting eq.\eqc{eq:Xvecdef} for $X^\mu$. Therefore, changing the covariant SSC to NW SSC does not change the three-point amplitude deduced from one-particle EFT.


\subsection{Minimal coupling in $s \gg 1$ limit as BHs}
As shown directly in~\cite{Guevara:2018wpp,Arkani-Hamed:2019ymq}, the classical-spin-limit of minimal coupling reproduces various classical observable of Kerr black holes, such as stress-tensor form factor and impulse. In the context of one-particle effective action, it was shown in~\cite{Chung:2018kqs} that the Wilson coefficients of minimal coupling at finite-spin deviate from that of Kerr black hole ($C_{\text{S}^n}=1$) by $\sim\frac{1}{s}$ terms. Thus in the limit $s\rightarrow \infty$ we recover Kerr black hole. Here we give a brief review of the map between the Wilson coefficient in the EFT basis and the coupling constants, $g_i$s, defined kinematically in the (anti)chiral spinor basis as follows. Begin with the general form of three-point amplitudes introduced in~\cite{Arkani-Hamed:2017jhn}:
\bl
\bld
M_s^{+2} &= \frac{\k m x^2}{2 m^{2s}} \left[ g_0 \la \bold{2} \bold{1} \ra^{2s} + g_1 \la \bold{2} \bold{1} \ra^{2s-1} \frac{x \la \bold{2} q \ra \la q \bold{1} \ra}{m} + \cdots + g_{2s} \frac{(x \la \bold{2} q \ra \la q \bold{1} \ra)^{2s}}{m^{2s}} \right] \,,
\\ M_s^{-2} &= \frac{\k m x^{-2}}{2 m^{2s}} \left[ g_0 [ \bold{2} \bold{1} ]^{2s} + g_1 [ \bold{2} \bold{1} ]^{2s-1} \frac{[ \bold{2} q ][ q \bold{1} ]}{xm} + \cdots + g_{2s} \frac{([ \bold{2} q ][ q \bold{1} ])^{2s}}{x^{2s}m^{2s}} \right] \,,
\eld \label{eq:3ptAnsatz}
\el
Here, we've expressed the coupling to the positive helicity graviton in the chiral spinor basis and the negative helicity graviton in the anti-chiral basis. For these choices, the minimal coupling simply corresponds to setting all couplings except $g_0$ to zero:
\bg
M_{s,min}^{+2} = \frac{\k m x^{2} }{2} \frac{\la \bf{21} \ra^{2s}}{m^{2s}} \,,\quad M_{s,min}^{-2} = \frac{\k m x^{-2} }{2} \frac{[ \bf{21} ]^{2s}}{m^{2s}}\,. \label{eq:3ptMinDef}
\eg
The minimal nature of the coupling can be seen in the high energy limit where all momenta are approximately massless, the expression matches to the minimal derivative three-point amplitude. Caution: the minimal coupling of eq.\eqc{eq:3ptMinDef} for $s>2$ is \emph{different} from the usual usage of minimal coupling in the QFT literature where the  derivatives of the kinetic term are simply covariantized, as shown in appendix \ref{app:3ptLag}.

In \cite{Chung:2018kqs} it was shown that at large $s$, the minimal couplings are matched to one-particle EFT with Wilson coefficients $C_{\text{S}^n} = 1 + \CO(1/s)$. To show this, we first work out the map between $C_{\text{S}^n}$ and $g_i$, by converting the EFT amplitude in eq.\eqc{EFT3pt}, into the chiral basis. This requires us to convert the square brackets to chiral spinors using the following two identities: 
\bl
\bld
[ \mathbf{21} ]  &=  \la \mathbf{21} \ra  + \frac{x \la \mathbf{2} q \ra \la q \mathbf{1} \ra}{m},\quad \frac{[ \mathbf{2} q ][q \mathbf{1} ]}{m x} &= - \frac{x \la \mathbf{2} q \ra \la q \mathbf{1} \ra}{ m} \,.
\eld \label{eq:conversionformula}
\el
We then arrive at:
\eqa
&&M_s^{+2} =\sum_{a+b\leq 2s}\; x^2 C_{\text{S}^{a+b}}  n^{s}_{a,b}\lambda^{2s}_2\left[ \mathbb{I}^{s-a} \left( - \frac{x |q \ra \la q|}{2m} \right)^a \left(\mathbb{I}  + \frac{x |q \ra \la q|}{m}\right)^{s-b} \left( - \frac{x | q \ra \la q | }{ 2m}  \right)^b\right]\lambda^{2s}_1\,.\nonumber\\
\eqae
Comparing with eq.\eqc{eq:3ptAnsatz} gives the following relation between $C_{\text{S}^n}$ and $g_i$
\bl
g_i &= \sum_{n=0}^{i} F_{i,n}^{s} C_{\text{S}^n},\quad F_{i,n}^{s} = \frac{1}{(-2)^n} \frac{(s!)^2}{(i-n)!} \sum_{m=0}^{n} \frac{1}{(s-m)!(s+m-i)!m!(n-m)!}\,. \label{eq:gfromCexact1}
\el
Equipped with this, we can derive the Wilson coefficients for minimal coupling, which sets $g_i=0$ for $i\neq0$. For example, since 
\eqa
g_0&=&C_{\text{S}^0},\quad g_1=s(C_{\text{S}^0}-C_{\text{S}^1}),\quad\nonumber\\
g_2&=&\frac{s^2(C_{\text{S}^2}{-}2C_{\text{S}^1}{+}C_{\text{S}^0})}{2}{+}\frac{s(2C_{\text{S}^1}{-}2C_{\text{S}^0}{-}C_{\text{S}^2})}{4}\,,
\eqae 
normalizing $g_0=1$, the vanishing of $g_1$ and $g_2$ sets
\bl
C_{\text{S}^2} = \frac{2s}{2s-1}\,. 
\label{eq:g2exact}
\el
This indeed tends to unity as one approaches the $s \gg 1$ limit. Similarly the vanishing of $g_3$ sets $C_{\text{S}^3} = \frac{2(s{+}1)}{2s{-}1}$. Thus in summary we see that the Wilson coefficients for minimally coupled particles deviate from that of Kerr black holes:
 \eq
 C^{Min,s}_{\text{S}^n}=C^{Kerr,s}_{\text{S}^n}{+}\mathcal{O}\left(\frac{1}{s}\right).
 \eqe 
It is not hard to work out the precise coefficients for $\frac{1}{s}$ corrections. A brief outline is given in appendix \ref{app:MinWilson}. Up to $\CO(s^{-2})$ order it can be shown analytically that
\bl\label{FiniteSpinWil}
C^{Min,s}_{\text{S}^{n}} &= 1 + \frac{n (n-1)}{4s} + \frac{(n^2-5n+10)n(n-1)}{32s^2} + \CO(s^{-3}) \,.
\el

\newpage
\section{Classical potential for general EFT at leading PN order } \label{sec:1PMPot}
\begin{figure}
\begin{center}
\includegraphics[scale=0.5]{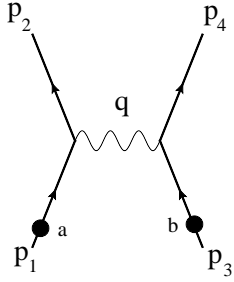}
\caption{The graviton exchange diagram between source $a$ and $b$ that yields the leading $\frac{1}{q^2}$ singularity, which is responsible for the classical potential.}
\label{GraviExchange}
\end{center}
\end{figure}

Here we derive the classical potential for general one-particle EFT to all orders in spin and leading PN for each spin degree. Here by classical, we are referring to the usual notion that the interaction is occurring at distances much greater than the de Broglie wavelength of its individual constituents. We begin with the graviton exchange diagram of spinning particles in the centre of momentum (COM) frame for the $2\rightarrow 2$ process. The kinematic set up is shown in fig.\ref{GraviExchange} and given by
\eqa
p_1 = ( E_a, \vec{p} {+} \vec{q} / 2 ), \quad p_3 = ( E_b, {-} \vec{p}{-} \vec{q} / 2 ),\quad  p_2 = ( E_a, \vec{p} {-} \vec{q} / 2 ) ,\quad  p_4 = ( E_b, {-} \vec{p} {+} \vec{q} / 2 ) 
\eqae
where the exchanged momentum $q^\mu=(p_{1}-p_2)^\mu=(0,\vec{q})$ is space-like, and on-shell conditions imply $\vec{p} \cdot \vec{q} = 0$. 
We also adopt the definitions $\vec{p} + \vec{q}/2 = \vec{p}_1 = - \vec{p}_3$, which will become useful when writing the potential. In the context of our $2 \to 2$ scattering, the distance between scattering bodies is associated with the impact parameter $|\vec{b}|$, where $\vec{b}$ is the Fourier transform of $\vec{q}$. Thus the classical limit is naturally associated with $|\vec{q}| \ll |\vec{p}|, m_a, m_b$, and $|\vec{J}|\sim |p|/|q|\gg1$, where $\vec{J}$ is the spatial part of the angular momentum~\cite{Cheung:2018wkq}. This hierarchy can be naturally matched to counting in Planck's constant $\hbar$, where we restore~\cite{Kosower:2018adc} (maintaining $c=1$)
\eqa\label{hbarA}
q^\m \to \hbar q^\m ,\;\;
 S^\m \to  \frac{1}{\hbar} S^\m,\;\; G \to \frac{1}{\hbar} G,\;\; m_{a,b} \to m_{a,b}, \;\;p^\m \to p^\m\,.
\eqae
The counting for $S^\mu$ is inherited from $J^\mu$, and $q^\m$ becomes the wavenumber. Thus in terms of $\hbar$ counting, the following combination is ``dimensionless" and contributes to the classical potential to all orders,
\eq
G|q|,\quad q S\,. 
\eqe
where we've suppressed the Lorentz indices for now. In particular, the expansion in $G|q|$ indicates that loop-level computations contains classical pieces, where one has 
\eq
 \frac{G}{q^2}\left(A+B G|q|+CG^2 q^2 \log q^2+\mathcal{O}(G^3)\right)
\eqe 
where $A,B,C$ are kinematic factors arising from tree, one-loop and two-loop computations respectively. This representation corresponds to the  Post-Minkowskian (PM) expansion. To make contact with the post-Newtonian (PN) expansion, one further expands in $\frac{p^2}{m^2}$, which correspond to the non-relativistic limit.

When the gravitating object is spinning, then $A,B,C$ also has dependence on the spin operator $S^\mu$. From the previous discussion, we have the following as the only linear in spin combinations with $\mathcal{O}(\hbar^0)$ scaling.
\eq
 q\cdot S,\quad \e(p_1,p_3,q,S)\equiv\e_{\m\n\l\s} p_1^\m p_3^\n q^\l S^\s = (E_a {+}E_b)\vec{p} \times \vec{q} \cdot \vec{S}\,.
 \eqe 
Note that due to the following identity for the product of two Levi-Civita tensors~\cite{vanOldenborgh:1989wn}, 
\bl
\bld
\e_{ijk} \e_{lmn} &= \left| \begin{array}{ccc} \delta_{il} & \delta_{im} & \delta_{in} \\ \delta_{jl} & \delta_{jm} & \delta_{jn} \\ \delta_{kl} & \delta_{km} & \delta_{kn} \end{array} \right|
\\ &= \delta_{il} (\delta_{jm}\delta_{kn} - \delta_{jn}\delta_{km}) - \delta_{im} (\delta_{jl}\delta_{kn} - \delta_{jn}\delta_{kl}) + \delta_{in} (\delta_{jl}\delta_{km} - \delta_{jm}\delta_{kl})\,, \nn
\eld
\el
any term that is even order in $\vec{p} \times \vec{q} \cdot \vec{S} $ can be successively reduced to a sum of polynomials of $p^2$ and $(q\cdot S)$, up to terms proportional to $q^2$ which do not contribute to the classical potential at leading perturbation order. Therefore the most general ansatz is at most linear in $\vec{p} \times \vec{q} \cdot \vec{S}$. Based on this power counting argument we expect the potential to have the following schematic form at leading order in Newton's constant $G$,
\bl
\bld
V(p,q,S) &= \frac{4 \pi G m_a m_b}{q^2} \left( F_0 + F_a \e(p_1,p_3,q,S_a) {+} F_b \e(p_1,p_3,q,S_b) \right)\,,
\eld \label{eq:1PMPotAnsatz}
\el
where $F_i$ are functions of  $p^2$, $m_{a,b}$, and $S_{a,b}\cdot q$. Note that while the spin operators are non-commuting, 
the commutator of spins $[S^i,S^j] = i \hbar \e^{ijk} S^k$ has an extra power of $\hbar$, so the difference between different orderings do not contribute at classical level. Finally
, each power of $S^\mu$ formally counts as 0.5 PN in the PN expansion because spin contains a factor of $1/c$ when restoring $c$.

Since we are interested in leading order in $G$ effects, the relevant information is contained in the exchange of single graviton between two sources, where the cubic coupling is given by the operators in the one-particle EFT. For spinning sources, there is a subtlety in obtaining the potential from the amplitude:  the external states form irreps of distinct Little groups. In other words, the amplitude is a matrix element in tensor product of distinct little group space, while the potential is understood to be matrix elements in the same Hilbert space.\footnote{Classical limits are really applicable to matrix elements and not operators. JWK would like to thank Nima Arkani-Hamed for pointing this out.} Thus in extracting the potential for spinning sources one needs to introduce a mapping procedure which we term  \emph{Hilbert space matching} (H.M.), which will be discussed in more detail in sections \ref{sec:HilbertMatch} and \ref{sec:FinSpin}.

As discussed previously, classical means that we are expanding in small $|q|$. In Lorentzian signature, $q^\mu$ is spacelike and $|q|\rightarrow 0$ translate to the zero momentum limit. On the other hand, by analytically continuing to complex (or split signature) momenta, we can have $|q|\rightarrow0$ correspond to null momenta, $q^2=0$. The advantage of such analytic continuation is that with $q^2=0$ the amplitude factorizes, with its residue given by the product of two three-point amplitudes. This approach was coined the \emph{holomorphic classical limit} (HCL) introduced by Guevara~\cite{Guevara:2017csg}, where the leading order potential is extracted as:
\eq
V(p,q)=\left.\frac{M_4(s,q^2)}{4 E_aE_b}\right|_{q^2\rightarrow0}=\frac{M_3\otimes M_3}{4 E_aE_b\,q^2}+{\rm H.M.}\,,
\eqe 
The residue is given by the product of three-point amplitudes, which we have computed for general EFTs in the previous section, and  the on-shell momenta for external particles are parameterized as:
\eqa\label{eq:4ptKinInput}
p_1 &=& \sket{\hat\eta} \bra{\hat\l} {+} \sket{\hat\l}\bra{\hat\eta},\;\; p_2 = \b' \sket{\hat\eta} \bra{\hat\l} {+} \frac{1}{\b'} \sket{\hat\l}\bra{\hat\eta} {+} \sket{\hat\l} \bra{\hat\l},\nonumber\\
p_3  &=& \sket{\eta} \bra{\l} {+} \sket{\l}\bra{\eta},\;\; p_4  = \b \sket{\eta} \bra{\l} {+} \frac{1}{\b} \sket{\l}\bra{\eta} {+} \sket{\l} \bra{\l}\,.
\eqae 
Note that each momentum is complex and the spinors are  constrained by $\la \hat\l \hat\eta \ra = [\hat\l \hat\eta] = m_a$ and $\la \l \eta \ra = [\l \eta] = m_b$. Note that since the transverse momentum is,
\bl
\bld
q^\mu = (p_1 - p_2)^\mu = (0,\vec{q}) &= - \sket{\hat\l}\bra{\hat\l} - (\b'-1) \left( \sket{\hat\eta}\bra{\hat\l} - \frac{1}{\b'} \sket{\hat\l}\bra{\hat\eta} \right) 
\\ &= \sket{\l}\bra{\l} + (\b-1)\left( \sket{\eta}\bra{\l} - \frac{1}{\b} \sket{\l}\bra{\eta} \right) \,,
\eld
\el
the parameters $\b$ and $\b'$ are required to satisfy the relations
\bl
q^2 &= - \frac{(\b'-1)^2 m_a^2}{\b'} = - \frac{(\b-1)^2 m_b^2}{\b} \,,
\el
and the HCL corresponds to taking $\beta,\beta'\rightarrow1$. 

Since in the HCL limit $q$ is null, the four-particle kinematics reduces to a product of two copies of three-particle kinematics, where we have:
\bl
\bld
\la \mathbf{21} \ra &= - \left( [ \mathbf{21} ] + \frac{[ \mathbf{2} \hat\l ][ \hat\l \mathbf{1} ]}{x_1 m_a} \right),\quad \frac{x_1 \la \mathbf{2} \hat\l \ra \la \hat\l \mathbf{1} \ra}{m_a} &= - \left( - \frac{[ \mathbf{2} \hat\l ][ \hat\l \mathbf{1} ]}{x_1 m_a} \right)
\\ \la \mathbf{43} \ra &= - \left( [ \mathbf{43} ] + \frac{[ \mathbf{4} \l ][ \l \mathbf{3} ]}{x_3 m_b} \right),\quad \frac{x_3 \la \mathbf{4} \l \ra \la \l \mathbf{3} \ra}{m_b} &= - \left( - \frac{[ \mathbf{4} \l ][ \l \mathbf{3} ]}{x_3 m_b} \right)
\eld \label{eq:HCL3ptkin}
\el
Here, the $x$-factors can be defined via $x_1 = \frac{\sbra{-\hat{\lambda}} p_1 \ket{\z}}{m_a \la -\hat{\lambda} \z \ra}$ and $x_3 = \frac{\sbra{\hat{\lambda}} p_3 \ket{\z}}{m_b \la \hat{\lambda} \z \ra}$. The product of $x$-factors can be expressed as
\bg
\bgd\label{xfactors}
\frac{x_1}{x_3} = \frac{u}{m_a m_b} = \r + \sqrt{\r^2 - 1},\quad\frac{x_3}{x_1}  = \frac{v}{m_a m_b} = \r - \sqrt{\r^2 - 1} .
\egd
\eg
where the variables $u$ and $v$ are defined as:\footnote{These variables satisfy the following useful identities
\bg
\bgd
\sbra{\eta} p_1 \ket{\eta} \sbra{\l} p_1 \ket{\l} = u v - m_a^2 m_b^2
\\ \sbra{\l} p_1 \ket{\l} = - \frac{(\b-1)^2}{\b} m_b^2 + (1-\b) v + \frac{\b-1}{\b} u 
\egd \label{eq:uvdef}\,.
\eg
}
\bl
\bld
u = \sbra{\l} p_1 \ket{\eta},\quad v = \sbra{\eta} p_1 \ket{\l}\,, \quad \r= \frac{p_1 \cdot p_3}{m_a m_b} = \frac{u + v}{2 m_a m_b}\,. \label{eq:rdef}
\eld
\el
Note that in the non-relativistic limit we have 
\eq
\rho=1{+}\frac{|\vec{p}|^2}{2\mu^2}{+}\mathcal{O}\left(\frac{|\vec{p}|^4}{\mu^4}\right)\,,
\eqe
where $\mu=\frac{m_am_b}{m_a+m_b}$ is the reduced mass, and the PN expansion correspond to expanding around $\r = 1$. Finally,  in the HCL limit, the operators $\e(p_1,p_3, q, S_a)$ and $q \cdot S_a$ are proportional to each other up to factors of $\sqrt{\rho^2-1}$: 
\bl
\bld
\e(p_1,p_3,q,S_a) &\;\underrightarrow{\rm \;\;\; HCL\;\;\;}  - i m_a^2 m_b \sqrt{\rho^2-1} \left( \frac{q \cdot S_a}{m_a} \right)  \,.
\eld \label{eq:LorInvOpBasisList}
\el
Thus in practice these two operators are disentangled by keeping track of which order in $\sqrt{\rho^2-1}$ they appear.

\subsection{The leading singularity and its PN expansion}
Starting with the three-point amplitude of general EFTs computed in sec.\ref{sec:EFT3pt}, eq.(\ref{eq:1bd3ptAmp}), we are now ready to construct the classical potential at 1 PM to all orders in spin. Gluing the three-point amplitudes and summing over the two intermediate helicity states one has:
\bl
\bld
\text{Res}_t &= A_{3a}^+ A_{3b}^- + A_{3a}^- A_{3b}^+
\\ &= (-1)^{s_a + s_b} \sum_{i=0}^{2s_a} \sum_{j=0}^{2s_b} \a^2 m_a^2 m_b^2 \frac{C_{\text{S}_a^i} C_{\text{S}_b^j}}{i! j!} \left( \frac{x^2_1}{ x_3^2} \left[ \ve^\ast(\bf{2}) \left( \frac{q \cdot S_a}{m_a} \right)^i \ve(\bf{1}) \right] \left[ \ve^\ast(\bf{4}) \left( \frac{q \cdot S_b}{m_b} \right)^j \ve(\bf{3}) \right] \right.
\\ &\phantom{=asdfasdfasdfasdf} + \left.   \frac{x^2_3}{ x_1^2} \left[ \ve^\ast(\bf{2}) \left( - \frac{q \cdot S_a}{m_a} \right)^i \ve(\bf{1}) \right] \left[ \ve^\ast(\bf{4}) \left( - \frac{q \cdot S_b}{m_b} \right)^j \ve(\bf{3}) \right]\right)
\\ &= (-1)^{s_a + s_b} \sum_{i=0}^{2s_a} \sum_{j=0}^{2s_b} B_{i,j} \left[ \ve^\ast (\mathbf{2}) \left( \frac{q \cdot S_a}{m_a} \right)^i \ve(\mathbf{1}) \right] \left[ \ve^\ast(\mathbf{4}) \left( - \frac{q \cdot S_b}{m_b} \right)^j \ve(\mathbf{3}) \right]\label{DefPol}
\eld
\el
The coupling constant $\a$ is defined as $\a = \k/2 = \sqrt{8 \pi G}$. The spins $s_a$ and $s_b$ are assumed to be integers. The sign factor $(-1)^{s_a + s_b}$ appears due to the choice of mostly minus metric signature; $\abs{\e(\bf{P})}^2 = (-1)^s$. This sign factor is irrelevant when computing the classical potential. The $B_{i,j}$ coefficient can be straight forwardly worked out to be
\bl
B_{i,j} &= \frac{(-1)^j  \frac{x^2_1}{ x_3^2} + (-1)^i  \frac{x^2_3}{ x_1^2}}{i! j!} \a^2 m_a^2 m_b^2 C_{\text{S}_a^i} C_{\text{S}_b^j} \,. \label{eq:1PMBcoeffs}
\el
One then simply substitute eq.(\ref{xfactors}) for the $x$-factor ratios and one obtains the full 1 PM result. For simplicity we will perform a PN expansion and keeping the leading PN result for each spin.\footnote{For full 1 PM result see~\cite{Chung:2020rrz}.} Taking the non-relativistic limit, the above yields:
\bl
\left. B_{i,j}\right|_{\rho\rightarrow 1} &= \frac{[(-1)^i + (-1)^j] - 2 \sqrt{\rho^2{-}1} [(-1)^i - (-1)^j]}{i! j!} \a^2 m_a^2 m_b^2 C_{\text{S}^i_a} C_{\text{S}^j_b}+\mathcal{O}(\rho^2{-}1) \,. \label{eq:1PMLPNBcoeffs}
\el
We are keeping the factor $\sqrt{\rho^2{-}1}$ since for $i{+}j=odd$ the leading term vanishes. We will proceed by stripping off the polarization tensors first, and consider the form of the remaining operator. The effects of the polarization tensors will manifest itself in the Hilbert space matching to be done later.

When $i + j$ is even, the first term is the leading PN contribution. When $i + j$ is odd, $\sqrt{\rho^2{-}1}$ term is the leading PN contribution. The two cases are treated separately.

\begin{itemize}
  \item $i + j$ even: $(-1)^i = (-1)^j$ can be used to simplify the expression.
\bl
\frac{B_{i,j}}{t} \left( \frac{q \cdot S_a}{m_a} \right)^i \left( - \frac{q \cdot S_b}{m_b} \right)^j &= - (-1)^{\frac{i+j}{2}} \frac{2 \a^2 m_a^2 m_b^2}{i! j! q^2} C_{\text{S}^i_a} C_{\text{S}^j_b} \left( \frac{- i \vec{q} \cdot \vec{S}_a}{m_a} \right)^i \left( \frac{- i \vec{q} \cdot \vec{S}_b}{m_b} \right)^j
\el
In position space, the expression becomes
\bl
- \frac{(-1)^{\frac{i+j}{2}} \a^2 m_a^2 m_b^2}{2 \pi i! j!} C_{\text{S}^i_a} C_{\text{S}^j_b} \left( \frac{\vec{S}_a}{m_a} \cdot \vec{\nabla} \right)^i \left( \frac{\vec{S}_b}{m_b} \cdot \vec{\nabla} \right)^j \frac{1}{r} \,.
\el
With non-relativistic flux normalisation $\frac{1}{4 E_a E_b} \simeq \frac{1}{4m_a m_b}$,
\bl
- \frac{(-1)^{\frac{i+j}{2}} C_{\text{S}^i_a} C_{\text{S}^j_b}}{i! j!} \left( \frac{\vec{S}_a}{m_a} \cdot \vec{\nabla} \right)^i \left( \frac{\vec{S}_b}{m_b} \cdot \vec{\nabla} \right)^j \frac{G m_a m_b}{r} \,.
\el

  \item $i + j$ odd: Due to the following vector identity~\cite{Chung:2018kqs}, any of $S_a$ or $S_b$ can be converted to spin-orbit coupling term.
\bl
\left[ \vec{v} \times \vec{S}_a \cdot \vec{q} \right] \left[ \vec{S}_b \cdot \vec{q} \right] = \left[ \vec{S}_a \cdot \vec{q} \right] \left[ \vec{v} \times \vec{S}_b \cdot \vec{q} \right] + q^2 \vec{v} \cdot \vec{S}_a \times \vec{S}_b
\el
The $q^2$ dependent part will combine with $q^2$ of the denominator to yield $q^0$ order expression, which does not contribute to long-distance effects. Therefore, there is a freedom for choosing which of $S_a$ or $S_b$ acquires spin-orbit coupling. The convention we choose is to attach spin-orbit coupling to odd powered spin to comply with the ansatz eq.\eqc{eq:1PMPotAnsatz}. For convenience, let us treat the cases separately.

For odd $i$ and even $j$, attach spin-orbit factor to $S_a$. The expression $\frac{B_{i,j}}{t} \left( \frac{q \cdot S_a}{m_a} \right)^i \left( - \frac{q \cdot S_b}{m_b} \right)^j$ is then evaluated as follows.
\bl
(-1)^{\frac{i+j+1}{2}} \frac{4 \a^2 m_a^2 m_b^2}{i! j! q^2} C_{\text{S}^i_a} C_{\text{S}^j_b} \left[ \left( \frac{\vec{p}_1}{m_a} - \frac{\vec{p}_3}{m_b} \right) \times \frac{\vec{S}_a}{m_a} \cdot (- i \vec{q}) \right] \left( \frac{- i \vec{q} \cdot \vec{S}_a}{m_a} \right)^{i-1} \left( \frac{- i \vec{q} \cdot \vec{S}_b}{m_b} \right)^j
\el
Going to position space and including non-relativistic flux normalisation factors, the contribution is evaluated as follows.
\bl
- \frac{2 (-1)^{\frac{i+j+1}{2}} C_{\text{S}^i_a} C_{\text{S}^j_b}}{i! j!}  \left[ \left( \frac{\vec{p}_1}{m_a} - \frac{\vec{p}_3}{m_b} \right) \times \frac{\vec{S}_a}{m_a} \cdot \vec{\nabla} \right] \left( \frac{\vec{S}_a}{m_a} \cdot \vec{\nabla} \right)^{i-1} \left( \frac{\vec{S}_b}{m_b} \cdot \vec{\nabla} \right)^j \frac{G m_a m_b}{r}
\el

For even $i$ and odd $j$, attach spin-orbit factor to $S_b$. The expression $\frac{B_{i,j}}{t} \left( \frac{q \cdot S_a}{m_a} \right)^i \left( - \frac{q \cdot S_b}{m_b} \right)^j$ is then evaluated as follows.
\bl
(-1)^{\frac{i+j+1}{2}} \frac{4 \a^2 m_a^2 m_b^2}{i! j! q^2} C_{\text{S}^i_a} C_{\text{S}^j_b} \left[ \left( \frac{\vec{p}_1}{m_a} - \frac{\vec{p}_3}{m_b} \right) \times \frac{\vec{S}_b}{m_b} \cdot (- i \vec{q}) \right] \left( \frac{- i \vec{q} \cdot \vec{S}_a}{m_a} \right)^{i} \left( \frac{- i \vec{q} \cdot \vec{S}_b}{m_b} \right)^{j-1}
\el
Going to position space and including non-relativistic flux normalisation factors, the contribution becomes the following.
\bl
\frac{2 (-1)^{\frac{i+j+1}{2}} C_{\text{S}^i_a} C_{\text{S}^j_b}}{i! j!}  \left[ \left( \frac{\vec{p}_1}{m_a} - \frac{\vec{p}_3}{m_b} \right) \times \frac{\vec{S}_b}{m_b} \cdot \vec{\nabla} \right] \left( \frac{\vec{S}_a}{m_a} \cdot \vec{\nabla} \right)^{i} \left( \frac{\vec{S}_b}{m_b} \cdot \vec{\nabla} \right)^{j-1} \frac{G m_a m_b}{r}
\el

\end{itemize}
Now that we have determined the polarization tensor-stripped part of the operator \eqc{DefPol}, we will proceed and account for the spin effects in the polarization tensors.  

\subsection{Hilbert space matching} \label{sec:HilbertMatch}
Scattering amplitudes are matrix elements between states that form irreps under distinct little groups, one for each particle. When considering our $2\rightarrow 2$ scattering, in fig.\ref{GraviExchange}, while states $1$ and $2$ are associated with the same massive particle $a$, their little group space is distinct since their momenta are different (the same goes with $3$ and $4$). As one can apply a Lorentz boost to transform between different momenta, the same transformation will relate the different little group spaces. This mapping of little group space will be termed~\textit{Hilbert space matching}. Since the polarization tensors form a natural little group basis, they are related by these Lorentz boosts and thus contain extra spin operators. The purpose of this section is to derive these spin factors rigorously. Note that a similar discussion was addressed for classical-spin variables of GR in~\cite{Levi:2015msa}\footnote{The authors would like to thank Mich\`ele Levi for informing us about the analogue.}.

Without loss of generality, we will consider the spin effects for polarization vectors. Let's begin with a reference momenta $p_0$ at rest. The corresponding polarization vector then takes the form
\eq
\ve^{\mu}_i (p_0)=\delta^{\mu}_i
\eqe
where the little group index on the polarization vector is aligned with the spatial directions. As we will see, having such a reference momenta allows us to define the little group frame in a uniform fashion. For the two body system, the natural reference momenta would be the momenta of the COM, i.e. 
\bl
p_{0,a/b}^\m &= \frac{m_{a/b}}{\sqrt{(p_1 {+} p_3)^2}} (p_1 {+} p_3)^\m \label{eq:RefMomDef}
\el
where $p_{0,a}$  $p_{0,b}$ are the COM momentum appropriately normalized such that it squares to $m_a^2$ and $m_b^2$ respectively. Now the polarization vector for generic momentum $p$ can be obtained by applying the boost that transforms $p_0$ to $p$, i.e. $G(p;p_0)^{\mu}\,_{\nu}$, and\footnote{ The explicit form of $G(p;p_0)^\m_{~\n}$ is given as:
\bl
\bld
G(p;p_0)^\m_{~\n} &= \delta^\m_{~\n} - \frac{(p +p_0)^\m (p + p_0)_{\n}}{(p \cdot p_0) + m^2} + \frac{2 p^\m p_{0 \n}}{m^2} \,.
\eld \label{eq:BoostExplicit}
\el
}
\bl
p^\m = G(p;p_0)^\m_{~\n} p_0^\n \,, \quad \ve_I^\m(p) = G(p;p_0)^\m_{~\n} \ve^\n_I (p_0) \,. \label{eq:BoostDef}
\el
Using this, we can relate the polarization vectors between in- and out-momenta, 
\bl\label{Relate1}
\ve_I (p_{out}) &= G(p_{out};p_0) G(p_{in};p_0)^{-1} \ve_I (p_{in})\,.
\el
Recall that the three-point amplitude, which will serve as the seed for the 1 PM potential, derived from our computation is understood as:
\eq\label{MAP}
\ve^{\ast \m}_I (p_{out})\, \mathcal{O}^K\,_J \,\ve_{K\m} (p_{in})\,,
\eqe
for the EFT operators $ \mathcal{O}^K\,_J$ that acts on the little group space.\footnote{For general QFT amplitudes operators act on Lorentz indices of the polarisation tensors. A systematic method to convert these operators to act on little group space will be given in section~\ref{sec:SysPolSpin}.} Using eq.(\ref{MAP}) we have:
\eqa
\ve^{\ast \m}_I (p_{out})\, \mathcal{O}^K\,_J \,\ve_{K\m} (p_{in})&=&  \e^{\ast \m}_I (p_{in}) \left[ G(p_{in};p_0) G(p_{out};p_0)^{-1} \right] \mathcal{O}^K\,_J \,\ve_{K\m} (p_{in})\nonumber\\
&=&\ve^{\ast \m}_I (p_{in})\, \widetilde{\mathcal{O}}^K\,_J \,\ve_{K\m} (p_{in}) \,,\label{eq:PolContIn}
\eqae
where $\widetilde{\mathcal{O}}^K\,_J$ is the matrix element acting on the little group space of the $in$-state particle. 

To extract the spin-operators in the Lorentz transformation, we first write:
\bl
\bld
G(p_A;p_B) &= e^{- i \l(p_A;p_B) p_A^\m p_B^\n J_{\m\n} }
\\ \l(p_A;p_B) &= \frac{\log \left[ \frac{1}{m^2} \left( p_A \cdot p_B + \sqrt{(p_A \cdot p_B)^2 - m^4} \right) \right] }{\sqrt{(p_A \cdot p_B)^2 - m^4}} \,.
\eld \label{eq:LorInvBoostDef}
\el
Next, we decompose $J^{\m\n}$ into ``rotation" and ``boost" part, denoted as $J^{\m\n}_{r}$ and $J^{\m\n}_{b}$ respectively. In the rest frame, $J^{\m\n}_{r}$ will be related to the spin-vector via,  $J^{\m\n}_{r} = - \frac{1}{m} \e^{\m\n\l\s} p_\l S_\s $. Thus a covariant definition of $J^{\m\n}_{r}$ would be the part of $J^{\m\n}$ which vanishes when acting on $p_{in}$, which is the frame our final operators are defined in. This indeed reduces to spatial rotations in the rest frame of $p_{in}$. Thus we have,
\eq
J^{\m\n}_{r} (p) = J^{\m\n} - \frac{1}{m^2} \left( p^\m p_\l J^{\l\n} + J^{\m\l} p_\l p^\n \right),\quad J^{\m\n}_{b} (p) = \frac{1}{m^2} \left( p^\m _\l J^{\l\n} + J^{\m\l} p_\l p^\n \right) \,.
\label{eq:LorGenDecomp}
\eqe
Now expanding out the combination of Lorentz generators relating the $out$ polarization vector to the $in$, we have 
\bl
\bld
\log \left[ G(p_{in};p_0) G(p_{out};p_0)^{-1} \right] &= - i p_{in}^\m \left( \l_{in} \left[1 - \frac{\l_{out} p_0 \cdot p_{out}}{2} \right] p_0^\n + \frac{\l_{in} \l_{out} m^2}{2} p_{out}^\n \right) J_{\m\n}
\\ &\phantom{=+} - i \l_{out} \left[ 1 - \frac{\l_{in} p_0 \cdot p_{in} }{2} \right] p_0^\m p_{out}^\n J_{\m\n}
\eld
\el
where $\l_{in,out}=\l(p_{in,out};p_0)$, and we truncated to second order in $\l_{in,out}$. Higher order terms will be higher in the PN expansion. Recasting in the basis of $J_b$ and $J_r$ (or equivalently $p_{in}^\m J_{\m\n}$ and $J_r$), where we denote the coefficients as $\D \l^\n$ and $\a_{so}$;
\bl
\bld
\log \left[ G(p_{in};p_0) G(p_{out};p_0)^{-1} \right] &= - i p_{in}^\m \D \l^\n J_{\m\n} - i {\a_{so} } p_0^\m p_{out}^\n J_{r,\m\n}
\\ &= - i p_{in}^\m \D \l^\n J_{\m\n} + i \frac{\a_{so}}{m} p_0^\m p_{out}^\n p_{in}^\l S_{in}^\s \e_{\m\n\l\s} \,.
\eld
\el
The spin $S_{in}^\m$ is defined via momenta  $p_{in}^\m$. In the non-relativistic limit we may take $\l_{in} \simeq \l_{out} \simeq m^{-2}$, leaving
\bg
\D \l^\n \simeq \frac{p_{out}^\n}{m^2} \,,\quad \a_{so} \simeq \frac{1}{2 m^2} 
\\ \log \left[ G(p_{in};p_0) G(p_{out};p_0)^{-1} \right] \simeq - i \frac{1}{m^2} p_{in}^\m p_{out}^\n J_{\m\n} + \frac{i}{2m^3} \e_{\m\n\l\s} p_0^\m p_{out}^\n p_{in}^\l S_{in}^\s \,. \label{eq:HilbertMatchLog}
\eg
This separation illustrates a very important difference: the second term \emph{depends} on the choice of reference momenta $p_0$ and are thus frame dependent while the first term is independent. The second term can be regarded as $\log \left[ G(p_{in};p_0) G(p_{out};p_0)^{-1} G(p_{out};p_{in}) \right]$ and generates spin-orbit interaction, and when we substitute $p_0$ defined in eq.\eqc{eq:RefMomDef} into the expression, we obtain
\bl
\frac{i}{2m^3} \e_{\m\n\l\s} p_0^\m p_{out}^\n p_{in}^\l S_{in}^\s = - \frac{i}{2m^2} \vec{p}_{out} \times \vec{p}_{in} \cdot \vec{S}_{in} = - \half \frac{\vec{p}}{m} \times \frac{\vec{S}_{in}}{m} \cdot ( - i \vec{q}) \,. \label{eq:HilbertMatchSO}
\el
The effects of the boost term $i p_{in}^\m \D \l^\n J_{\m\n} = \log \left[ G(p_{out};p_{in})^{-1} \right]$, which is \emph{independent} of reference momentum $p_0$, are suppressed in powers of $1/s$ as we will show in section~\ref{sec:FinSpin}, so they can be neglected in the infinite spin limit.

In summary, the presence of contracted polarization tensors leads to 
\bl
\e^\ast (p_2) \cdot \e(p_1) &= \e^\ast(p_1) \left[ \iden - \left( \frac{\vec{p}_1}{m_a} \times \frac{\vec{S}_a}{m_a} \right) \cdot \frac{- i \vec{q}}{2} + \cdots \right] \e(p_1) \,. \label{eq:HilMatA}
\el
For particle $b$, there is an additional sign factor due to definition of $\vec{q}$, which is consistent with the dictionary provided in \cite{Holstein:2008sx}.
\bl
\e^\ast (p_4) \cdot \e(p_3) &= \e^\ast(p_3) \left[ \iden + \left( \frac{\vec{p}_3}{m_b} \times \frac{\vec{S}_b}{m_b} \right) \cdot \frac{- i \vec{q}}{2} + \cdots \right] \e(p_3) \,. \label{eq:HilMatB}
\el
In sum, the overall effect is to multiply all the results obtained in the previous sections by the factor
\bl
1 - \half \left[ \frac{\vec{p}_1}{m_a} \times \frac{\vec{S}_a}{m_a} - \frac{\vec{p}_3}{m_b} \times \frac{\vec{S}_b}{m_b} \right] \cdot (- i \vec{q}) = 1 - \half \left[ \frac{\vec{p}_1}{m_a} \times \frac{\vec{S}_a}{m_a} - \frac{\vec{p}_3}{m_b} \times \frac{\vec{S}_b}{m_b} \right] \cdot \vec{\nabla} \label{eq:HilbertMatch}
\el
and truncating to leading PN order. This effect can be compared to $V_{\text{kin}}$ of eq.(3.32) in \cite{Vines:2016qwa}, which is an augmentation of $V_{\text{el}} + V_{\text{mag}}$ in eq.(3.31) by the factor $\left[ 1 - \half \left( \mathbf{v}_1 \times \mathbf{a}_1 - \mathbf{v}_2 \times \mathbf{a}_2 \right) \right]$ followed by truncation to leading PN order.

In the above we have chosen to define our potential as an operator acting on the little group space of the incoming particle. One could alternatively choose that of the outgoing particle, but the result will not change since the Hilbert-space matching factor relevant for $s \to \infty$ limit is simply a Lorentz rotation, i.e. from $|out\rangle=R|in\rangle$ we have $\langle out|in\rangle=\langle in|R^T|in\rangle=\langle out|R^T|out\rangle$.  
\subsection{The general/Kerr 1 PM leading PN potential} \label{sec:1PMsummary}
Combining the terms arising from Hilbert space matching, the general leading PN all order in spin classical potential is given as
\bl
\bld
V_{cl} &= - \sum_{m,n=0}^{\infty} \frac{(-1)^{n} C_{\text{S}^{2n-m}_a} C_{\text{S}^m_b}}{(2n-m)! m!} \left( \frac{\vec{S}_a}{m_a} \cdot \vec{\nabla} \right)^{2n-m} \left( \frac{\vec{S}_b}{m_b} \cdot \vec{\nabla} \right)^m \frac{G m_a m_b}{r}
\\ &\phantom{=}- \sum_{m,n=0}^{\infty} \frac{2 (-1)^{m+n} C_{\text{S}^{2m+1}_a} C_{\text{S}^{2n}_b}}{(2m+1)! (2n)!}  \left[ \left( \frac{\vec{p}_1}{m_a} - \frac{\vec{p}_3}{m_b} \right) \times \frac{\vec{S}_a}{m_a} \cdot \vec{\nabla} \right] \left( \frac{\vec{S}_a}{m_a} \cdot \vec{\nabla} \right)^{2m} \left( \frac{\vec{S}_b}{m_b} \cdot \vec{\nabla} \right)^{2n} \frac{G m_a m_b}{r}
\\ &\phantom{=}- \sum_{m,n=0}^{\infty} \frac{2 (-1)^{m+n} C_{\text{S}^{2m}_a} C_{\text{S}^{2n+1}_b}}{(2m)! (2n+1)!}  \left[ \left( \frac{\vec{p}_1}{m_a} - \frac{\vec{p}_3}{m_b} \right) \times \frac{\vec{S}_b}{m_b} \cdot \vec{\nabla} \right] \left( \frac{\vec{S}_a}{m_a} \cdot \vec{\nabla} \right)^{2m} \left( \frac{\vec{S}_b}{m_b} \cdot \vec{\nabla} \right)^{2n} \frac{G m_a m_b}{r}
\\ &\phantom{=} + \sum_{m,n=0}^{\infty} \frac{(-1)^{n} C_{\text{S}^{2n-m}_a} C_{\text{S}^m_b}}{2(2n-m)! m!} \left( \left[ \frac{\vec{p}_1}{m_a} \times \frac{\vec{S}_a}{m_a} - \frac{\vec{p}_3}{m_b} \times \frac{\vec{S}_b}{m_b} \right] \cdot \vec{\nabla} \right) \left( \frac{\vec{S}_a}{m_a} \cdot \vec{\nabla} \right)^{2n-m} \left( \frac{\vec{S}_b}{m_b} \cdot \vec{\nabla} \right)^m \frac{G m_a m_b}{r}
\eld \label{eq:1PMClassPot}
\el
Up to quartic order in spin, the results match with the known results available in the literature~\cite{tulczyjew1959equations,Barker:1975ae,Hergt:2008jn,Levi:2010zu,Levi:2014gsa}.

For Kerr black holes, all Wilson coefficients are unity. The summation in the first line of the above can be simplified as follows:
\bl
- \sum_{n=0}^{\infty}  \frac{(-1)^n}{(2n)!} \left[ \left( \frac{\vec{S}_a}{m_a} + \frac{\vec{S}_b}{m_b} \right) \cdot \vec{\nabla} \right]^{2n} \frac{G m_a m_b}{r} \label{eq:evenpot}
\el
The above expression reproduces $- \cosh (\mathbf{a_0} \times \mathbf{\nabla}) \frac{m_1 m_2}{R}$ of eq.(3.31) in \cite{Vines:2016qwa}, where the following notation had been adopted.
\bl
(\vec{S} \cdot \vec{\nabla})^2 \frac{1}{r} &\equiv - (\vec{S} \times \vec{\nabla})^2 \frac{1}{r} \label{eq:VinesNot}
\el
The difference of both sides does not contribute to long-distance dynamics as it is some multiple of Dirac delta. Using the notation eq.\eqc{eq:VinesNot} on eq.\eqc{eq:evenpot} yields
\bl
- \sum_{n=0}^{\infty} \frac{1}{(2n)!} \left[ \left( \frac{\vec{S}_a}{m_a} + \frac{\vec{S}_b}{m_b} \right) \times \vec{\nabla} \right]^{2n} \frac{G m_a m_b}{r} &= - \cosh \left[ \left( \frac{\vec{S}_a}{m_a} + \frac{\vec{S}_b}{m_b} \right) \times \vec{\nabla} \right] \frac{G m_a m_b}{r} \,.
\el
For the second and third line, once again we can use eq.\eqc{eq:VinesNot} to simplify it to the form:
\bl
- \frac{2 C_{\text{S}^i_a} C_{\text{S}^j_b}}{i! j!}  \left[ \left( \frac{\vec{p}_1}{m_a} - \frac{\vec{p}_3}{m_b} \right) \cdot \frac{\vec{S}_b}{m_b} \times \vec{\nabla} \right] \left( \frac{\vec{S}_a}{m_a} \times \vec{\nabla} \right)^{i} \left( \frac{\vec{S}_b}{m_b} \times \vec{\nabla} \right)^{j-1} \frac{G m_a m_b}{r} \,.
\el
Setting all Wilson coefficients to unity gives:
\bl
- \sum_{n=0}^{\infty} \frac{2}{(2n+1)!} \left( \frac{\vec{p}_1}{m_a} - \frac{\vec{p}_3}{m_b} \right) \cdot \left[ \left( \frac{\vec{S}_a}{m_a} + \frac{\vec{S}_b}{m_b} \right) \times \vec{\nabla} \right]^{2n+1} \frac{G m_a m_b}{r}
\el
which can be formally written as
\bl
- 2 \left( \frac{\vec{p}_1}{m_a} - \frac{\vec{p}_3}{m_b} \right) \cdot \sinh \left[ \left( \frac{\vec{S}_a}{m_a} + \frac{\vec{S}_b}{m_b} \right) \times \vec{\nabla} \right] \frac{G m_a m_b}{r} \,.
\el
This expression matches $ - 2 (\mathbf{v}_1 - \mathbf{v}_2) \cdot \sinh (\mathbf{a_0} \times \mathbf{\nabla}) \frac{m_1 m_2}{R}$ of eq.(3.31) in \cite{Vines:2016qwa}. Applying similar identities to the last line we find the complete 1 PM leading PN potential for rotating black holes: 
\bl
\bld
V_{cl}^{\text{BBN}} &= \left( - \cosh \left[ \left( \frac{\vec{S}_a}{m_a} + \frac{\vec{S}_b}{m_b} \right) \times \vec{\nabla} \right] - 2 \left( \frac{\vec{p}_1}{m_a} - \frac{\vec{p}_3}{m_b} \right) \cdot \sinh \left[ \left( \frac{\vec{S}_a}{m_a} + \frac{\vec{S}_b}{m_b} \right) \times \vec{\nabla} \right] \right) \frac{G m_a m_b}{r}
\\ &\phantom{=} + \half \left( \left[ \frac{\vec{p}_1}{m_a} \times \frac{\vec{S}_a}{m_a} - \frac{\vec{p}_3}{m_b} \times \frac{\vec{S}_b}{m_b} \right] \cdot \vec{\nabla} \right) \cosh \left[ \left( \frac{\vec{S}_a}{m_a} + \frac{\vec{S}_b}{m_b} \right) \times \vec{\nabla} \right] \frac{G m_a m_b}{r}\,.
\label{eq:1PMClassBHPot}
\eld
\el
The above result can be matched to eq.(3.31) and eq.(3.32) in \cite{Vines:2016qwa}. The first few terms of eq.\eqc{eq:1PMClassBHPot} are; 
\bl
\bld
V_{cl} &= - \frac{G m_a m_b}{r} + \frac{G}{r^2} \hat{n} \cdot \left[ \frac{4m_a + 3m_b}{2 m_a} \vec{p}_1 \times \vec{S}_a - (a \leftrightarrow b) \right]
\\ &\phantom{=} - \frac{G}{r^3} \left( \delta_{i,j} - 3 \hat{n}_i \hat{n}_j \right) \left[ \frac{m_b}{2 m_a} C_{\text{S}_a^2} S_a^i S_a^j + (a \leftrightarrow b) + S_a^i S_b^j \right]
\eld \label{eq:BHPotEx}
\el
where $\hat{n} = \frac{\vec{r}}{r}$.

We end this section with a curious observation; the classical potential eq.\eqc{eq:1PMClassPot} mod the Hilbert space matching terms in eq.\eqc{eq:HilbertMatch}, matches to intermediate results in EFT computations in~\cite{Porto:2005ac, Levi:2010zu, Levi:2014gsa} where higher time derivatives has not yet been eliminated. In EFT computations the classical potential is obtained by fixing spin variable gauges and doing Feynman diagram computations. The result obtained at this point will in general contain higher time derivatives such as $\dot{\vec{S}}$ and $\dot{\vec{v}}$, and neglecting these higher time derivatives will give an expression equivalent to eq.\eqc{eq:1PMClassPot} without the last line; for example, compare the results given in this section with eq.(48) of \cite{Porto:2005ac} and eq.(71) of \cite{Levi:2010zu} for spin-orbit interactions, terms proportional to $C_{1(\text{BS}^3)}$ in eq.(3.10) of \cite{Levi:2014gsa} for $(S_a)^3$ interactions, and the sum of first two terms proportional to $C_{1(\text{ES}^2)}$ in eq.(3.10) of \cite{Levi:2014gsa} for $(S_a)^2 S_b$ interactions. The last procedure of EFT computations is eliminating higher time derivatives through redefinition of variables. Since the last line of eq.\eqc{eq:1PMClassPot} is generated through Hilbert space matching procedure, this procedure generates the terms corresponding to terms generated from redefining variables to eliminate higher time derivatives.


\section{Classical potential from finite spins} \label{sec:FinSpin}
As was shown in previous sections, the Wilson coefficients $C_{\text{S}^n}$ that generate the amplitude for minimal coupling is unity plus $\mathcal{O}(s^{-1})$ deformations. Na\"ively, this would lead one to expect that the potential computed from minimal coupling at finite spin is distinct from that of Kerr black holes. However, as discussed earlier, in computing the potential one needs to perform Hilbert-space matching, which carries its own finite-spin effects. Remarkably, the corrections induced from the boost operator cancels the effects from the  $\mathcal{O}(s^{-1})$ deformations of the Wilson coefficients, rendering the ``effective" Wilson coefficient to be unity, and thus reproduces the correct Kerr black hole! Here we give a detailed discussion of such finite-spin cancellation, and comment on the some of the previous results in the literature. 
\subsection{Hilbert space matching at finite spins and minimal coupling as black holes} \label{sec:MinCoupBH}
Consider the following on-shell three-point kinematics where momentum $\vec{q}$ is complex null; $( \vec{q} )^2 = 0$. We will restrict to integer spins in this section.
\bg
p_1 = (m, \vec{0}) \,,\quad q = (0, \vec{q}) \,,\quad p_2 = (m, - \vec{q}) \label{eq:3ptKinConfig}
\eg
Recall that the finite-spin effects from the Hilbert-space matching are due to boost generators $\log \left[ G(p_{out};p_{in})^{-1} \right]$ in eq.\eqc{eq:HilbertMatchLog} and are independent of the choice of reference momentum $p_0$. Thus, for the purpose of discussing finite spin effects, we are at liberty to set $p_0$ of to $p_1$ and eliminate Thomas-Wigner rotation effects, i.e. the second term in \eqc{eq:HilbertMatchLog} vanishes. The little group matrix element $\ve^\ast_I (\bf{2}) \cdot \ve_J (\bf{1})$ is computed as
\bl
\bld
\ve^\ast_I (\bf{2}) \cdot \ve_J (\bf{1}) &= \ve^\ast_I (\bf{1}) \left[ e^{i \frac{\vec{q}}{m} \cdot \vec{K}} \right] \ve_J (\bf{1}) = \ve^\ast_I (\bf{1}) \left[ \sum_{n=0}^{s} \frac{(-1)^n}{(2n)!} \left( \frac{\vec{q} \cdot \vec{K} }{m} \right)^{2n} \right] \ve_J (\bf{1})
\\ &= \ve^\ast_I (\bf{1}) \left[ \sum_{n=0}^{s} \frac{(-1)^n (2s-2n)!}{(2s)!} {s \choose n} \left( \ \frac{{q} \cdot {S} }{m} \right)^{2n} \right] \ve_J (\bf{1})
\eld \label{eq:3ptBoost2Rot}
\el
where eq.\eqc{eq:Boost2Rot} has been used to obtain the last line together with the condition $( \vec{q} )^2 = 0$. The coefficient of $(q \cdot S)^{2n}$ scales as $(-4s)^{-n}$ in the limit $s \to \infty$, so they are finite spin effects for $n \neq 0$. Inserting these finite spin pieces into eq.\eqc{eq:1bd3ptAmp} will give the following result.
\bl
\bld
M_s^{2\eta} &= \frac{\k m x^{2 \eta}}{2} \ve^\ast_I (\bf{1}) \left[ \sum_{i=0}^{s} \frac{(-1)^i (2s-2i)!}{(2s)!} {s \choose i} \left( - \eta \frac{{q} \cdot {S} }{m} \right)^{2i} \right] \left[ \sum_{j=0}^{2s} \frac{C_{\text{S}^j}}{j!} \left( - \eta \frac{q \cdot S}{m} \right)^j \right] \ve_J (\bf{1})
\\ &= \frac{\k m x^{2 \eta}}{2} \ve^\ast_I (\bf{1}) \left[ \sum_{n=0}^{2s} \frac{C_{\text{S}^n_{eff}}}{n!} \left( - \eta \frac{q \cdot S}{m} \right)^n \right] \ve_J (\bf{1})\,,
\eld \label{eq:EFT3ptWCeff}
\el
where we've incorporated the Hilbert-space matching terms to define the effective Wilson coefficient $C_{\text{S}^n_{eff}}$. The relation between $C_{\text{S}^n}$ and $C_{\text{S}^n_{eff}}$ is given as:
\bl
\bld
\frac{C_{\text{S}^m_{eff}}}{m!} &= \sum_{i=0}^{\lfloor m/2 \rfloor} \frac{(-1)^i (2s-2i)!}{(2s)!} {s \choose i} \frac{C_{\text{S}^{m-2i}}}{(m-2i)!}
\\ &= \sum_{n=0} \left( \delta_{m,n} - \frac{\delta_{m-n,2}}{4s} + \frac{\delta_{m-n,4} - 4 \delta_{m-n,2}}{32s^2} + \CO(s^{-3}) \right) \frac{C_{\text{S}^n}}{n!}\,. \label{eq:bareWC2eff}
\eld
\el
One can interpret $C_{\text{S}^n_{eff}}$ as the $2^n$-multipole of the particle which would be measured by an observer at infinity.\footnote{Indeed the reference frame $p_0$ chosen here is very similar to the "body-fixed frame" introduced in~\cite{Levi:2015msa}.} Remarkably, substituting the Wilson coefficients for minimal coupling while keeping the finite-spin effects, for example eq.(\ref{FiniteSpinWil}), we find that the effective $C_{\text{S}^n_{eff}}$ turns out to be unity! In other words,

\noindent \begin{center} \framebox{ \parbox{0.95\textwidth}{Minimal coupling reproduces the Kerr Black hole Wilson coefficients at finite spins once the Hilbert space matching terms are included!  }} \end{center}
The complete proof of this statement is given in appendix \ref{sec:resrep}.

Note that since the boost part of the Hilbert space matching is independent of choice of $p_0$, the above statement would hold for the choice used to evaluate the potential, i.e. eq.\eqc{eq:RefMomDef}. Thus when combined with the spin-orbit part, we can conclude that minimal coupling reproduces the potential for Kerr black holes for finite $s$, in the sense that it reproduces the correct spin-dependent  terms up to degree $2s$ in spin operators.  In the following, we will verify the above statement directly on the amplitudes computed from Feynman rules. 
\subsection{Example: gravitational potential from spin-1 spin-0 scattering} \label{sec:SysPolSpin}
Let us consider a concrete example to verify that the potential derived from the scattering amplitude of spinning particles matches with the Kerr black hole potential, without resorting to taking the infinite-spin limit. We will use the scattering of massive spin-1 and scalar particle as an example. This system has been computed previously in~\cite{Holstein:2008sx} and~\cite{Vaidya:2014kza} with conflicting results. We will comment on the root of their discrepancy in the next section. As we will now show, indeed the correct potential up to degree two in spin-operator is reproduced.

The amplitude of a graviton exchange between a vector and a scalar is given by~\cite{Vaidya:2014kza}\footnote{We have simplified the expression and corrected the sign of $\eta_{\m\n}$. This expression is equivalent to eq.(70) of~\cite{Holstein:2008sx}.}
\bl
iM = - \frac{4\pi G m_a m_b}{\vec{q}^2} \varepsilon_{2}^{\ast\m} \left[ - \eta_{\m\n} + 2 \hat{p}_{3\m} \hat{p}_{2\n} + 2 \hat{p}_{1\m} \hat{p}_{3\n} - \hat{p}_{1\m} \hat{p}_{2\n} \right] \varepsilon_{1}^\n \,, \label{eq:ExPot}
\el
where the hatted variables are defined as $\hat{p}_i^\m = p_i^\m / m_i$ and $m_i$ are the masses of the particles respectively, with $m_1=m_2=m_a$ and $m_3=m_4=m_b$. We've suppressed the little group indices for simplicity. Note that unlike our previous expression, where the Lorentz indices of the polarization vectors are contracted with each other, which was the case since we were considering the factorization limit, here they also contract with the momenta. Thus schematically we have: 
\bl
\varepsilon_{2,\{\m_s\}}^\ast \,\CO^{\{\m_s\},\{\n_s\}}\, \varepsilon_{1,\{\n_s\}} = \varepsilon_{1,I,\{\l_s\}}^\ast \, [G(p_1;p_0) G(p_0;p_2)]^{\{\l_s\}}_{~~\{\m_s\}} \, \CO^{\{\m_s\},\{\n_s\}} \, \varepsilon_{1,\{\n_s\}} \,.
\el
We can extract the matrix elements through an appropriately generalised version of eq.\eqc{eq:OpAction}. Returning to the example eq.\eqc{eq:ExPot}, we need to express $\varepsilon^\ast_2$ using $\varepsilon^\ast_1$. In the non-relativistic limit the approximation $\hat{p}_i \cdot \hat{p}_j \simeq 1$ can be applied to eq.\eqc{eq:BoostExplicit}, yielding
\bl
\varepsilon^\ast_{2\m} &\simeq \varepsilon^\ast_{1\a} \left[ \delta^\a_\m - \half \hat{p}_{0}^\a \hat{p}_{1 \m} + \half (\hat{p}_{0}^\a \hat{p}_{2 \m} - \hat{p}_{2}^\a \hat{p}_{0 \m}) - \half \hat{p}_{2}^\a \hat{p}_{2 \m} \right] \,. \label{eq:pol2inpol1}
\el
Inserted into the numerator of eq.\eqc{eq:ExPot} then yields,
\bl
- \varepsilon^{\ast \m}_1 \left[ \eta_{\m\n} + \half ( \hat{p}_{0\m} \hat{p}_{2\n} - \hat{p}_{2\m} \hat{p}_{0\n} ) - 2 ( \hat{p}_{3\m} \hat{p}_{2\n} - \hat{p}_{2\m} \hat{p}_{3\n} ) + \frac{1}{2} \hat{p}_{2\m} \hat{p}_{2\n} \right] \varepsilon_1^\n \,. \label{eq:ExPotPol}
\el
We now recast the above expression in terms of spin operators. 

Recall that the spin operator matrix element in little group space $\mathbb{S}^\m$,  defined in eq.\eqc{eq:OpAction}, is given by
\bl
\mathbb{S}^\m(p) &= (-1) \ve_{\r}^\ast (p) \left( - \frac{1}{2m} \e^{\m\n\l\s} p_\n \left[ J_{\l\s} \right]^{\r}_{~\delta} \right) \ve^\delta (p) = \frac{i}{m} \e^{\m\n\l\s} p_\n \ve_{\l}^\ast (p) \ve_{\s} (p) \,,\label{eq:spin1spin}
\el
where one keeps in mind that the polarization vectors carry little group indices. An additional sign factor appears due to metric signature; $\ve_{\m}^\ast \ve^\m = - \mathbb{I}$.\footnote{ The matrix elements for the Lorentz generator $J^{\m\n}$ is given as:
\bl
\left[ J^{\m\n} \right]^\r_{~\s} &= i [ \eta^{\m\r} \delta^\n_\s - \eta^{\n\r} \delta^\m_\s ]\,.
\el}
The squared spin operator $(\mathbb{S}^\m \mathbb{S}^\n)$ is computed as
\bl
\bld
(\mathbb{S}^\m \mathbb{S}^\n) &= - \frac{1}{m^2} \e^{\m\a\b\g} \e^{\n\l\r\s} p_\a p_\l \ve^\ast_{\b} \ve_{\s} \sum_{\ell} \ve_{\ell,\g} \ve^\ast_{\ell,\r} = - \ve^\m \ve^{\ast \n} - \mathbb{I} \left( \eta^{\m\n} - \frac{p^\m p^\n}{m^2} \right)
\eld \label{eq:spin1spinspin}
\el
where we've used the completeness relation for summing over the little group index $\ell$:
\bl
\sum_{\ell} \ve^\m_{\ell} \ve^\ast_{\ell,\n} = - \delta^\m_{\n} + \frac{p^\m p_\n}{m^2} \,.
\el
Equations eq.\eqc{eq:spin1spin} and eq.\eqc{eq:spin1spinspin} can be rewritten in the following form respectively.
\bl
\ve_i^{\ast \m} \ve_j^\n - \ve_i^{\ast \n} \ve_j^\m &= - \frac{i}{m} \e^{\m\n\l\s} p_\l \mathbb{S}_{ij,\s}
\\ \ve_i^{\ast \m} \ve_j^\n + \ve_i^{\ast \n} \ve_j^\m &= - (\mathbb{S}^\m \mathbb{S}^\n + \mathbb{S}^\n \mathbb{S}^\m)_{ij} - 2 \delta_{ij} \left( \eta^{\m\n} - \frac{p^\m p^\n}{m^2} \right)
\el
Averaging over the two, we get
\bl
\bld
\ve_i^{\ast \m} \ve_j^\n &= - \delta_{ij} \left( \eta^{\m\n} - \frac{p^\m p^\n}{m^2} \right) - \frac{i}{2m} \e^{\m\n\l\s} p_\l \mathbb{S}_{ij,\s} - \half (\mathbb{S}^\m \mathbb{S}^\n + \mathbb{S}^\n \mathbb{S}^\m)_{ij}
\\ &= - \delta_{ij} \left( \frac{\eta^{\m\n}}{2} - \frac{p^\m p^\n}{m^2} \right) - \frac{i}{2m} \e^{\m\n\l\s} p_\l \mathbb{S}_{ij,\s} - \left(\frac{\mathbb{S}^\m \mathbb{S}^\n + \mathbb{S}^\n \mathbb{S}^\m}{2} - \frac{\eta^{\m\n} \mathbb{S}_\a \mathbb{S}^\a}{4} \right)_{ij}
\eld \label{eq:spin1polfull}
\el
The above expression can be used to convert amplitudes written in terms of polarisation vectors into potentials with spin variables. The second line is separated into trace/anti-symmetric/symmetric traceless terms and reflects the fact that $\eta_{\m\n}(S^\m S^\n)_{ij} = - 2 \delta_{ij}$ is a quadratic Casimir. Combining eq.\eqc{eq:ExPotPol} and eq.\eqc{eq:spin1polfull}, eq.\eqc{eq:ExPot} becomes
\bl
\bld
iM &\simeq - \frac{4\pi G}{\vec{q}^2} \left[ \iden + \frac{i}{2m_a^3} \e (p_0,p_2,p_1,\mathbb{S}) - \frac{2i}{m_a^2 m_b} \e (p_3,p_2,p_1,\mathbb{S}) + \frac{(q \cdot \mathbb{S})^2}{2m_a^2} \right]_{ij}
\\ &= - \frac{4\pi G}{\vec{q}^2} \left[ \iden - \frac{1}{2} \frac{\vec{p}_1}{m_a} \times \frac{\vec{\mathbb{S}}}{m_a} \cdot (- i \vec{q}) + \frac{2(m_a + m_b)}{m_b} \frac{\vec{p}_1}{m_a} \times \frac{\vec{\mathbb{S}}}{m_a} \cdot (- i \vec{q}) + \frac{(\vec{q} \cdot \vec{\mathbb{S}})^2}{2m_a^2} \right]_{ij}
\eld \label{eq:spin1spin0pot}
\el
for PN order of our interest, yielding the correct potential eq.\eqc{eq:BHPotEx} up to $S_a^2$ order.

\subsection{Comparison with existing literature} \label{sec:SpinCompare}
The treatment of polarisation tensors in the previous section is a generalisation of the treatment given in~\cite{Maybee:2019jus}, where $p_0$ was implicitly chosen to be equal to $p_{in}$. 
For example, the constrast of the treatment can be seen by comparing eq.(4.12) of~\cite{Maybee:2019jus}
\bl
\ve_i^{\ast \m} (p_1 + \hbar \bar{q}) \ve_j^\n (p_1) &= \ve_i^{\ast \m} \ve_j^\n - \frac{\hbar}{m_1^2} \left( \bar{q} \cdot \ve_i^\ast \right) p_1^\m \ve_j^\n - \frac{\hbar^2}{2 m_1^2} \left( \bar{q} \cdot \ve_i^\ast \right) \bar{q}^\m \ve_j^\n + \CO(\hbar^3) \,,
\el
to our eq.\eqc{eq:pol2inpol1}; the anti-symmetric combination $(\hat{p}_0 \hat{p}_2 - \hat{p}_2 \hat{p}_0)$ is missing in the above formula. The results are consistent since we recover the above formula from eq.\eqc{eq:pol2inpol1} as $p_0 \to p_1$.
This means the amplitude eq.(4.21) of~\cite{Maybee:2019jus} needs to be augmented by the corresponding spin-orbit factor, which can be implemented as the following transform.
\bl\label{Addon}
\delta^{ij} \to \delta^{ij} - \frac{i}{2m_1^3} \e_{\m\n\l\s} p_0^\m (\hbar \bar{q})^\n p_1^\l \left( {s^{ij}}/{\hbar} \right)^\s
\el
A sign difference has been introduced to compensate for difference in the conventions; the convention is $\e^{0123} = +1$ in this paper, while $\e_{0123} = +1$ in~\cite{Maybee:2019jus}. Incorporating this contribution into eq.(4.21) of~\cite{Maybee:2019jus} yields the following expression.
\bl
\bld
\hbar^3 \CM^{ij}_{1-0} &= - \left( \frac{\k}{2} \right)^2 \frac{4}{\bar{q}^2} \left[ \left( (p_1 \cdot p_2)^2 - \half m_1^2 m_2^2 \right) ( \delta^{ij} - \frac{i}{2m_1^3} \e_{\m\n\l\s} p_0^\m \bar{q}^\n p_1^\l s^{\s \, ij} )   \right.
\\ &\phantom{=} \left. - \frac{i}{m_1} (p_1 \cdot p_2) p_1^\r \bar{q}^\s p_2^\l \e_{\r\s\l\delta} s_1^{\delta \,ij} + \frac{1}{2m_1^2} \left( (p_1 \cdot p_2)^2 - \half m_1^2 m_2^2 \right) (\bar{q} \cdot s^{ik}_1)( \bar{q} \cdot s^{kj}_1 ) + \CO(\hbar^2) \right]
\eld \nn\\
\el
Going to the COM frame and taking $p_0$ to be at rest in this frame, the leading PN contribution is equivalent to eq.\eqc{eq:spin1spin0pot}. Note that the extra term introduced in eq.(\ref{Addon}) is a rotation factor which drops out in the polarization sum used to define $\langle \Delta s\rangle$ in~\cite{Maybee:2019jus}.  Thus our observation that the finite-spin approach, which once again entails keeping $s$ finite and absorbing an associated factor of $\hbar$, reproduces the classical-spin dynamics is in accordance with the matching observed in~\cite{Maybee:2019jus}.  

In~\cite{Holstein:2008sx,Vaidya:2014kza}, spatial components of polarisation tensors were simply considered to be equivalent to polarisation tensors of the reference momentum, which was taken to be at rest in the COM frame. Up to linear order in spin the na\"ive treatment will give the same answer, but starting at quadratic order in spin the results start to deviate. For example, eq.(77) of~\cite{Holstein:2008sx}
\bl
\e^{b\ast}_f \cdot \e_i^b \simeq - \hat{\e}^{b\ast}_f \cdot \hat{\e}^b_i + \frac{1}{m_b^2} \hat{\e}^{b\ast}_f \cdot \vec{p} \, \hat{\e}^b_i \cdot \vec{p} + \frac{i}{2 m_b^2} \vec{S}_b \cdot \vec{p} \times \vec{q} - \frac{1}{4 m_b^2} \hat{\e}^{b\ast}_f \cdot \vec{q} \, \hat{\e}^b_i \cdot \vec{q}
\el
is clearly incompatible with our eq.\eqc{eq:pol2inpol1} due to the factor of $(4m^2)^{-1}$ in $(\vec{\e} \cdot \vec{q})^2$. 
The same computation is listed as eq.(16) in~\cite{Vaidya:2014kza} and spin-quadratic term ($\vec{q} \cdot \hat{\e_1} \vec{q} \cdot \hat{\e_2^\ast}$) matches, but it is unclear how this result has been derived.
\bl
\e^\ast(p_2) \cdot \e(p_1) \approx - \hat{\e_1} \cdot \hat{\e_2^\ast} - \frac{1}{2 m_a^2} \vec{q} \cdot \hat{\e_1} \vec{q} \cdot \hat{\e_2^\ast} - \frac{1}{2 m_a^2} (q^i p^j - p^i q^j) \hat{\e_1}^i \hat{\e_2^\ast}^j
\el

\section{Universality and the classical-spin limit } \label{sec:1PMrevisit}
While the discussions so far revolves around amplitude representations using polarization tensors, the power of modern amplitude techniques can only be fully put in force when the expressions are given in purely on-shell variables. The simplicity of minimal coupling eq.\eqc{eq:3ptMinDef} compared to the EFT amplitude eq.\eqc{eq:1bd3ptAmp} is one manifestation of this fact. Another example is provided by the scattering amplitude for Compton scattering~\cite{Arkani-Hamed:2017jhn,Chung:2018kqs,Johansson:2019dnu};
\bl
\bld
A_4(p_1,k_2^{+1}, k_3^{-1}, p_4) &= \a^2 \frac{\bra{3} p_1 \sket{2}^{2}}{\bra{2} p_1 \sket{2} \bra{2} p_4 \sket{2}} \left( \frac{ [ \bold{1} 2 ] \la 3 \bold{4} \ra  + \la \bold{1} 3 \ra [ 2 \bold{4} ] }{\bra{3} p_1 \sket{2}} \right)^{2s}
\\ A_4(p_1,k_2^{+1}, k_3^{+1}, p_4) &= \a^2 \frac{{m^2}[23]^2}{\bra{2} p_1 \sket{2} \bra{2} p_4 \sket{2}} \frac{\la \bold{1} \bold{4} \ra^{2s}}{m^{2s}} \,.
\eld
\el
The simplicity of the three-point amplitude for minimal coupling is manifested when the external spinors are converted into preferred (anti-)chiral basis using the Dirac equation~\cite{Arkani-Hamed:2017jhn}. Since minimal coupling yields the dynamics of Kerr black holes, this motivates us to look into the problem of computing the classical potential of minimally coupled particles in the chiral basis. Remarkably, we observe that for the residue of the tree-level graviton exchange and the triangle coefficient for the one-loop two graviton exchange:

\noindent \begin{center} \framebox{ \parbox{0.95\textwidth}{ \textit{the coefficient of the spin operators of degree $i$ and $j$ in particle $a$ and $b$ respectively takes the form:
\bg\label{Universal}
\CA_{i,j}^{s_a,s_b} = A_{i,j} (-1)^{2s_a {+} 2s_b} \frac{(2s_a)!}{(2s_a - i)!} \frac{(2s_b)!}{(2s_b-j)!} \,,
\eg
where $s_a$ and $s_b$ is the spin of the two particles. Importantly as long as $A_{i,j}$ is non-vanishing, which requires $ i\leq 2s_a $ and $ j\leq 2s_b$, it is independent of spin. } }} \end{center}

\noindent In other words, the each term in the residue factorizes into a universal part and a spin-dependent part which is only comprised of combinatoric factors. Recall that the relation between minimal coupling and Kerr black hole is unambiguously established in the strict $s\rightarrow \infty$ limit~\cite{Arkani-Hamed:2019ymq}. The above identity allows us to access the asymptotic limit from finite spin computations! This is indeed the approach taken in~\cite{Chung:2018kqs}, which reproduced the classical potential up to quartic order in spin. 

Let us now demonstrate eq.\eqc{Universal}, by evaluating the $t$-channel residue eq.\eqc{DefPol} for minimal coupling in the anti-chiral basis:
\bl
\bld
\text{Res}_t &= A_{3a}^+ A_{3b}^- + A_{3a}^- A_{3b}^+
\\ &= \a^{2} m_a^2 m_b^2 \left\{ \frac{x^2_1}{x^2_3} (-1)^{2s_b} \la \bold{2} \bold{1} \ra^{2s_a} [ \bold{4} \bold{3} ]^{2s_b} + \frac{x^2_3}{x^2_1} (-1)^{2s_a} [\bold{2}  \bold{1} ]^{2s_a} \la \bold{4} \bold{3} \ra^{2s_b} \right\} \,. \label{eq:1PMAnsatzDemo1}
\eld
\el
The sign factors $(-1)^{2s_a}$ and $(-1)^{2s_b}$ are remnants of taking momenta $p_2$ and $p_4$ as outgoing. Again using eq.\eqc{eq:HCL3ptkin}, this can be converted into the anti-chiral basis
\bl
\bld
\label{eq:1PMAnsatzDemo2}
\text{Res}_t&=\sum_{i=0}^{2s_a} \sum_{j=0}^{2s_b} \CA_{i,j}^{s_a,s_b} \left( \sbra{\mathbf{2}}^{2s_a} \left(\frac{\sket{\hat\l}\sbra{\hat\l}}{x_1 m_a} \right)^i \sket{\mathbf{1}}^{2s_a} \right) \left( \sbra{\mathbf{4}}^{2s_b} \left( \frac{\sket{\l}\sbra{\l}}{x_3 m_b} \right)^j \sket{\mathbf{3}}^{2s_b} \right)\\
&=\sum_{i=0}^{2s_a} \sum_{j=0}^{2s_b} \CA_{i,j}^{s_a,s_b} x^i y^j\\
 \CA_{i,j}^{s_a,s_b}&=  \;\a^2 m_a^2 m_b^2 \left( \frac{x^2_1}{x^2_3} \frac{ (2s_a)! \delta_{j,0}}{(2s_a - i)! i!} +\frac{x^2_3}{x^2_1}  \frac{ \delta_{i,0} (2s_b)!}{(2s_b-j)! j!} \right)
\eld
\el
Note that only $\CA_{i,0}$ and $\CA_{0,j}$ are non-vanishing. Importantly the only dependence of $\CA_{i,j}^{s_a,s_b}$ on spins of external particles $s_a$ and $s_b$ is through combinatoric factors. In other words, defining:
\bg
\CA_{i,j}^{s_a,s_b} = A_{i,j} N^{s_a,s_b}_{i,j} \,, \quad N^{s_a,s_b}_{i,j} = (-1)^{2s_a + 2s_b} \frac{(2s_a)!}{(2s_a - i)!} \frac{(2s_b)!}{(2s_b-j)!} \,. \label{eq:universality1}
\eg
the factors $A_{i,j}$\footnote{The $A_{i,j}$ coefficient here is the same as the $\tilde{A}_{i,j}$ in~\cite{Chung:2018kqs}} are \emph{independent} of spins $s_a$ and $s_b$, proving the relation eq.\eqc{Universal}. Note that for generic values of non-zero $g_{i>0}$ the above property will not hold! Here, in the leading PN expansion, $ A_{i,j}=\a^2 m_a^2 m_b^2\left( \frac{ \delta_{j,0}}{i!} + \frac{ \delta_{i,0}}{ j!} \right)$. As we will demonstrate in appendix \ref{sec:2PMUniv}, the same universality behaviour occurs at one-loop up to $s=2$ where minimal coupling leads to unique Compton amplitude.

To obtain the classical potential, we simply take the classical-spin limit. We introduce formal parameters $\tilde{x} = 2s_a x$ and $\tilde{y} = 2s_b y$ and hold $\tilde{x},\tilde{y}$ fixed while taking the $s_a, s_b \to \infty$ limit, yielding
\bl
\lim_{s_a, s_b \to \infty} \text{Res}_t &= \sum_{i,j=0}^{\infty} A_{i,j} \tilde{x}^i \tilde{y}^j \,, \label{eq:AcoeffAnsatz1}
\el
where integer spins are assumed. We can now compare with the polarization tensor basis to obtain the spin-dependent pieces of the potential. Recall that $t$-channel residue is given in the last line of eq.(\ref{DefPol}), which we recast into on-shell matrix elements using eq.\eqc{PSDef}. As in eq.\eqc{DefPol} we assume the spins $s_a$ and $s_b$ to be integers.
\bl
\bld
Res_t &= (-1)^{s_a + s_b} \sum_{i=0}^{2 s_a} \sum_{j=0}^{2s_b} B_{i,j} \sum_{k=0}^{i} \tilde{n}^{s_a}_{i,k} \la \mathbf{21} \ra^{s_a - k} \left( - \frac{x_1 \la \mathbf{2} \hat\l \ra \la \hat\l \mathbf{1} \ra}{m_a} \right)^{k} [\mathbf{21}]^{s_a - i + k} \left( \frac{[ \mathbf{2} \hat\l ][ \hat\l \mathbf{1} ]}{x_1 m_a} \right)^{i - k}
\\ &\phantom{=asdfasdfasdf} \times \sum_{l=0}^{j} \tilde{n}^{s_b}_{j,l} \la \mathbf{43} \ra^{s_b - l} \left( - \frac{x_3 \la \mathbf{4} \l \ra \la \l \mathbf{3} \ra}{m_b} \right)^{l} [\mathbf{43}]^{s_b - j + l} \left( \frac{[ \mathbf{4} \l ][ \l \mathbf{3} ]}{x_3 m_b} \right)^{j - l} \,.\label{eq:SpinInSpinorVar}
\eld
\el
The coefficients $\tilde{n}^{s}_{i,k}$ are rescaled version of $n^s_{i,j}$ coefficient in eq.\eqc{EFT3pt}.
\bl
\tilde{n}^{s}_{i,k} &= \frac{i!}{2^i} { s \choose k} {s \choose i-k } \,.
\el
Using eq.\eqc{eq:HCL3ptkin} to convert all angle brackets to square brackets and recasting in terms of $\mathbb{I}$, $x$, and $y$ variables as before results in the following expression.
\bl
Res_t &= \sum_{i=0}^{2 s_a} \sum_{j=0}^{2s_b} \sum_{k=0}^{i} \sum_{l=0}^{j} B_{i,j} \tilde{n}^{s_a}_{i,k} \tilde{n}^{s_b}_{j,l} \left(\mathbb{I} + x \right)^{s_a - k} x^{i} \left(\mathbb{I} + y \right)^{s_b - l} y^{j} \,. \label{eq:BcoeffFinSpin}
\el
The classical limit proceeds as the previous case; we introduce formal parameters $\tilde{x} = 2s_a x$ and $\tilde{y} = 2s_b y$, and then take $s_a, s_b \to \infty$ limit while holding $\tilde{x}$ and $\tilde{y}$ finite. 
\bl
\lim_{s_a, s_b \to \infty} Res_t &= \sum_{i,j=0}^{\infty} \frac{B^{\infty}_{i,j}}{2^{i+j}} \tilde{x}^i \tilde{y}^j e^{\tilde{x}/2 + \tilde{y}/2} \label{eq:BcoeffAnsatz1}
\el
Matching eq.\eqc{eq:BcoeffAnsatz1} to eq.\eqc{eq:AcoeffAnsatz1} gives a solution for the coefficient of minimal coupling, $B_{i,j}^{min,\infty}$, as a function of $A_{i,j}$:
\eq\label{eq:BcoeffSol1}
\framebox[6cm][c]{$\displaystyle B_{i,j}^{min, \infty}  = 2^{i+j} \sum_{k,l} \frac{A_{i-k,j-l}}{(-2)^{k+l} k! l!}$} 
\eqe
As one can see the classical potential, computed from $B_{i,j}^{min, \infty}$, can be computed from $A_{i,j}$ which are spin independent. In other words due to the universal behaviour of minimal coupling, \textit{allows us to obtain the classical-spin limit from simple manipulation of finite-spin amplitudes}! Augmented with the Hilbert space matching term, gives the full potential. Indeed the RHS of eq.(\ref{eq:BcoeffSol1}) is what was used to compute the potential in~\cite{Chung:2018kqs}. 

While demonstration of universality has been restricted to tree level computations, the same property also holds at one-loop level as shown in appendix \ref{sec:2PMUniv}. 
The origin of this universality reflects the fact that the minimal coupling three-point amplitude for general spins can be viewed as a one parameter family with a special feature: \textit{the 3-pt amplitude of spin-$s$ is simply $2s$ power of that of spin-$\frac{1}{2}$}. 
We remark that it is possible to extend this combinatoric structure to non-minimal couplings as shown in appedix \ref{sec:resrep}.


\section{Conclusions} \label{sec:conc}
In this paper, we derive the spin-dependent pieces of the leading post-Newtonian order gravitational potential for general spinning body from the one-particle effective action, to all orders in spin. We first cast the EFT operators into three-point amplitudes, parameterized by the EFT Wilson coefficient. Special care was taken in defining the spin-operators, which acts on the physical Hilbert space, i.e. the irreps of massive Little group. This requires us to map the polarization vectors to a common basis, which for the two-body problem we choose to be the center of mass momenta, generating extra spin-operator dependent terms. We refer to this procedure as Hilbert-space matching. After gluing the three-point amplitudes and with the addition of the Hilbert space matching terms, we derive the leading post-Newtonian order classical potential for general spinning objects, to all orders in spin. As consistency checks, we compare the result to known quartic order in spin results for general spinning compact bodies, and then compare to the known equivalent order potential for binary black holes
. We stress the importance of choosing an appropriate reference momentum to define the basis of the Little group space, as the spin states of the gravitating bodies should be labelled consistently throughout the stellar binary evolution and thus so must the reference momentum.

Our result was derived by going to the classical-spin limit, i.e. taking $s\rightarrow \infty$. At finite spins, it is known that the Wilson coefficients for minimal coupling deviates from unity by finite spin effects. On the other hand, it is clear that the Hilbert space matching also has finite spin effects. Thus it is interesting to see what the potential looks like for minimal coupling at finite spins. Remarkably, we find that minimal coupling at finite spin reproduces the black hole potential to the prescribed spin order! In other words, the finite spin effects cancel each other. 
Whether such cancellation is a feature of current perturbation order or a feature that continues to higher perturbation orders remains as a problem to be explored. We also comment on the various discrepancies in earlier work on classical-spinning potentials from the scattering of spin-$\frac{1}{2}$ and $1$ particles~\cite{Holstein:2008sx, Vaidya:2014kza}.

Finally, we consider the computation of the classical potential using amplitudes in the uniform chirality basis, where the massive spinors of the external legs are converted to the same chirality. This basis is more ``natural'' for the amplitudes as expressions simplifies. When restricted to minimal coupling in the HCL, we find that for finite spins, the potential factorizes into a spin-dependent combinatoric factor and a spin-independent function. We term such factorization as universality, reflecting the fact that the spin-independent function is universal for all spins. This allows us to compute this factor using finite spins, and then simply replace the combinatoric factors with its classical-spin limit. This yields the formula first proposed in~\cite{Chung:2018kqs}, which was argued intuitively. 


While finite spin particles are expected to reproduce dynamics of Kerr BHs when taking into account the boost effects suppressed in powers of $\frac{1}{s}$, this statement is only true for the polarisation tensor basis. For example, the boost effects in anti-chiral basis is not necessarily suppressed in powers of $\frac{1}{s}$ as is evident from eq.\eqc{eq:MinKerr}. However, we may use universality to take the classical-spin limit in the anti-chiral basis and separate the boost effects. Considering that expressions using polarisation tensors are likely to be unavailable as we go to higher loops, universality is expected to be a useful tool to directly access the classical-spin limit and compute the potentials.

\section{Acknowledgements}
The authors would like to thank Nima Arkani-Hamed, Sangmin Lee, Mich\`ele Levi, Alexander Ochirov, Rafael Porto and Justin Vines for helpful discussions.
 MZC and YTH is supported by MoST Grant No. 106-2628-M-002-012-MY3. YTH is also supported by Golden Jade fellowship. The work of JWK  was supported in part by the National Research Foundation of Korea (NRF) Grant 2016R1D1A1B03935179. JWK was also supported in part by Kwanjeong Educational Foundation.

\appendix

\section{Lagrangian description of minimal coupling} \label{app:3ptLag}
For $s\leq2$, the minimal coupling eq.\eqc{eq:3ptMinDef} coincides with the minimal coupling for Dirac fermion, $F^{\mu\nu}F_{\mu\nu}$ for vectors, and Rarita-Schwinger for spin-$\frac{3}{2}$ and KK gravity for spin-2. These reflect the fact that in the high-energy limit where the kinematics becomes massless, the amplitude becomes the minimal coupling for self-interacting fields. Beyond spin-2 no such construction is known in flat-space. In conventional QFT literature minimal coupling is simply the covariantization of the kinetic terms, which is simply the d'Alembertian operator plus terms that enforces the state to be transverse traceless. The goal of this appendix is to explicitly demonstrate that our higher spin minimal coupling do not match with such case, where the action is constructed from the viewpoint of Weinberg's textbook~\cite{Weinberg:1995mt}; the free kinetic action is the inversion of the propagator. The conventions for spinor-helicity variables are the same as the ones used in \cite{Chung:2018kqs} throughout the appendices.

The kinetic term for arbitrary integer spin field that is responsible for the factor $\frac{1}{p^2 - m^2}$ in the propagator can be promoted to curved space as follows.
\bl
S_{kin} &= \int \sqrt{-g} \frac{(-1)^{s}}{2} \left( D^{\m} \phi^{\n_1 \cdots \n_s} D_{\m} \phi_{\n_1 \cdots \n_s} - m^2 \phi^{\n_1 \cdots \n_s} \phi_{\n_1 \cdots \n_s} \right)
\el
The sign factor $(-1)^s$ is there to make sure that the kinetic term for physical degrees of freedom have the right sign. 
%
This action contributes to the three-point amplitude through the derivative $D$, and the contribution in the chiral (undotted) basis is given as follows.
\bl
M_3 &= \frac{\k x^2}{2 m^{2s-2}} \sum_{i=0}^{2s} g_i^{\text{kin}} \la \bold{2} \bold{1} \ra^{2s-i} \left( \frac{x \la \bold{2}3 \ra \la 3\bold{1} \ra}{m} \right)^i
\\ g_i^{\text{kin}} &= { s \choose i } - s { s-1 \choose i-1 } = - (i - 1) { s \choose i }
\el
%
The propagator also contains projection terms in the numerator which project onto physical degrees of freedom. Terms in the Lagrangian that are responsible for such projection terms on flat space will in general contain terms where derivatives on the fields are contracted to one of the indices of the fields that derivatives act on.
\bl
(\p_{\m} \phi^{\m}_{~\n_1 \cdots \n_{s-1} }) (\p_{\l} \phi^{\l \n_{1} \cdots \n_{s-1}}) \,.
\el
Doing integration by parts, this term can be cast as follows.
\bl
(\p_{\m} \phi^{\l}_{~\n_1 \cdots \n_{s-1} }) (\p_{\l} \phi^{\m \n_{1} \cdots \n_{s-1}}) \,.
\el
In the above expression, derivatives are contracted to the field that it does not act on. These expressions can be promoted to curved space as follows.
\bl
(\p_{\m} \phi^{\m}_{~\n_1 \cdots \n_{s-1} }) (\p_{\l} \phi^{\l \n_{1} \cdots \n_{s-1}}) &\Longrightarrow g^{\a\b} g^{\m\n} g^{\l_1 \s_1} \cdots g^{\l_{s-1} \s_{s-1}} (D_{\a} \phi_{\b \l_1 \cdots \l_{s-1}}) (D_{\m} \phi_{\n \s_1 \cdots \s_{s-1}}) \label{eq:expr1}
\\ (\p_{\m} \phi^{\l}_{~\n_1 \cdots \n_{s-1} }) (\p_{\l} \phi^{\m \n_{1} \cdots \n_{s-1}}) &\Longrightarrow g^{\a\b} g^{\m\n} g^{\l_1 \s_1} \cdots g^{\l_{s-1} \s_{s-1}} (D_{\a} \phi_{\m \l_1 \cdots \l_{s-1}}) (D_{\n} \phi_{\b \s_1 \cdots \s_{s-1}}) \label{eq:expr2}
\el
Up to surface terms, eq.\eqc{eq:expr1} and eq.\eqc{eq:expr2} differ by a term linear in the curvature tensor $R_{\m\n} = [D_\m , D_\n]$. Therefore, any expression of the form eq.\eqc{eq:expr2} can be converted to the form eq.\eqc{eq:expr1} by introducing extra Riemann tensor couplings.
\bl
\bld
\sqrt{-g} (g^{\a\b} g^{\m\n} - g^{\a\n}g^{\b\m}) g^{\l_1 \s_1} \cdots g^{\l_{s-1} \s_{s-1}} (D_{\a} \phi_{\b \l_1 \cdots \l_{s-1}}) (D_{\m} \phi_{\n \s_1 \cdots \s_{s-1}})
\\ \propto \p [\cdots] + \sqrt{-g} g^{\a\b} g^{\m\n} g^{\l_1 \s_1} \cdots g^{\l_{s-1} \s_{s-1}} (\phi_{\b \l_1 \cdots \l_{s-1}}) ( [D_{\a} , D_{\m}] \phi_{\n \s_1 \cdots \s_{s-1}})
\eld \label{eq:KinTermDefAmb}
\el
Linear coupling to $h$ obtained from the substitution $g^{\m\n} \to \eta^{\m\n} - \k h^{\m\n}$ and $D_\m \to \p_\m + \G_\m$ on eq.\eqc{eq:expr1} will not contribute to on-shell three-point amplitude, due to transverse nature of on-shell physical DOF; $p_\m \ve(p)^\m = 0$. Also, terms linear in the curvature tensor cannot affect $g_1$ and only can affect $g_{i \ge 2}$. This shows $g_1 = 0$ is a constraint that cannot be changed for coupling to gravitons. 
To remove the ambiguity coming from eq.\eqc{eq:KinTermDefAmb}, expression of the form eq.\eqc{eq:expr1} and its generalisation to multiple derivatives will be considered as the canonical expression for terms introduced to kill unphysical degrees of freedom. 

The following coupling of the curvature tensor to higher-spin fields generate electric couplings.
\bl
\CL^{\text{ES}^{2j-2}}_{int} = m^{4-2j} \left( \p_{\m_1} \cdots \p_{\m_{j-2}} \p_{\n_1} \cdots \p_{\n_{j-2}} R_{\m_{j-1}\n_{j-1}\m_{j}\n_{j}} \right) \phi^{\m_1 \cdots \m_{j} \s_{1} \cdots \s_{s-j}} \phi^{\n_1 \cdots \n_{j}}_{\phantom{\n_1 \cdots \n_{j}}\s_{1} \cdots \s_{s-j}} \label{eq:ElecCoup1}
\el
Although in principle covariant derivatives $D$ must be used, considering them as partial derivatives $\p$ suffices for analysing 3pt amplitudes. 
This piece is related to the Wilson coefficient $C_{\text{ES}^{2j-2}}$ appearing in one-particle effective action for point particles. The contribution of this coupling to the 3pt amplitude in the chiral (undotted) basis is given below.
\bl
\CL^{\text{ES}^{2n}}_{int} &= \frac{(\overbrace{\p \cdots \p}^{2n-2} R) \phi \phi}{m^{2n-2}} \to \frac{\k x^2}{2 m^{2s-2}} \sum_{i=2n}^{2s} g^{\text{ES}^{2n}}_i \la \bold{2} \bold{1} \ra^{2s-i} \left( \frac{x \la \bold{2}3 \ra \la 3\bold{1} \ra}{m} \right)^i \label{eq:EScoup1Res}
\\ g^{\text{ES}^{2n}}_i &= - \frac{(-1)^{s} (-1)^{n}}{2^{n-1}} { s - n - 1 \choose i - 2n } \,,\, n \leq s-1
\el
As $n$ is restricted to the range $n \leq s-1$ for eq.\eqc{eq:ElecCoup1}, the coupling that affects $C_{\text{ES}^{2s}}$ needs to be introduced independently. The following coupling will do the job, but it turns out that this coupling is unnecessary.
\bl
\left( \p_{\m_1} \cdots \p_{\m_{s-1}} \p_{\n_1} \cdots \p_{\n_{s-1}} R_{\m_{s}\n_{s+1}\m_{s+1}\n_{s}} \right) \left( \p^{\m_{s+1}} \phi^{\m_1 \cdots \m_{s}} \right) \left( \p^{\n_{s+1}} \phi^{\n_1 \cdots \n_{s}} \right)
\el

The magnetic couplings are generated by the following couping to the curvature tensor.
\bl
\CL^{\text{BS}^{2j-1}}_{int} = m^{2-2j} \left( \p_{\m_1} \cdots \p_{\m_{j-2}} \p_{\n_1} \cdots \p_{\n_{j-1}} R_{\m_{j-1}\n_{j}\m_{j}\n_{j+1}} \right) \phi^{\m_1 \cdots \m_{j} \s_{1} \cdots \s_{s-j}} \left( \p^{\n_{j+1}} \phi^{\n_1 \cdots \n_{j}}_{\phantom{\n_1 \cdots \n_{j}}\s_{1} \cdots \s_{s-j}} \right)
\el
This piece is related to the Wilson coefficient $C_{\text{BS}^{2j-1}}$
, and this coupling's contribution to the 3pt amplitude in the chiral (undotted) basis becomes the following.
\bl
\CL^{\text{BS}^{2n+1}}_{int} &= \frac{(\overbrace{\p \cdots \p}^{2n-1} R) \phi (\p \phi)}{m^{2n}} \to \frac{\k x^2}{2 m^{2s-2}} \sum_{i=2n+1}^{2s} g^{\text{BS}^{2n+1}}_i \la \bold{2} \bold{1} \ra^{2s-i} \left( \frac{x \la \bold{2}3 \ra \la 3\bold{1} \ra}{m} \right)^i \label{eq:BScoup1Res}
\\ g^{\text{BS}^{2n+1}}_i &= \frac{(-1)^{s} (-1)^{n}}{2^{n}} { s - n - 1 \choose i - 2n - 1 } \,,\, n \leq s-1
\el
All magnetic couplings up to $C_{\text{BS}^{2s-1}}$ are covered by this coupling.

We empirically find that defining the following coefficients
\bl
{c}_{\text{ES}^{2n}} &= (-1)^s (-2)^{n-1} \frac{(s+n-1)\cdots(s-n)}{(2n)!} = (-1)^s (-2)^{n-1} { s + n - 1 \choose 2n } \label{eq:cESanswer}
\\ 
{c}_{\text{BS}^{2n+1}} &= - (-1)^s (-2)^{n} \frac{(s+n-1)\cdots(s-n-1)}{(2n+1)!} = - (-1)^s (-2)^{n} { s + n - 1 \choose 2n + 1 } \label{eq:cBSanswer}
\el
gives the following sum
\bl
\bld
g_{i}^{\text{min}} &= g_{i}^{\text{kin}} + \sum_{n=0}^{s-1} ( g^{\text{ES}^{2n}}_{i} c_{\text{ES}^{2n}} + g^{\text{BS}^{2n+1}}_{i} c_{\text{BS}^{2n+1}} )
\\ &= - (i-1) { s \choose i } + \sum_{n=1}^{s-1} \left[ { s - n - 1 \choose i - 2n }{ s + n - 1 \choose 2n } - { s - n - 1 \choose i - 2n - 1 }{ s + n - 1 \choose 2n + 1 } \right]
\\ &= \delta_{i,0}
\eld
\el
which holds for $0 \leq i \leq 2s$\footnote{The sum has been checked up to $s=200$ numerically.}.  
In other words, the following action
\bl
S_{min} = S_{kin} + \int d^4x \left[ c_{\text{ES}^{2n}} \CL^{\text{ES}^{2n}}_{int} + c_{\text{BS}^{2n+1}} \CL^{\text{BS}^{2n+1}}_{int} \right]
\el
corresponds to minimal coupling of higher-spin fields to gravitons.

\section{Wilson coefficients for minimal coupling} \label{app:MinWilson}
To compute $\frac{1}{s}$ corrections to Wilson coefficients for minimal coupling, the inverse matrix of the matrix $F^s_{i,n}$ in eq.\eqc{eq:gfromCexact1} is needed. $F^s_{i,n}$ can be expanded as an asymptotic series in $\frac{1}{s}$;
\bl
\bld
F^{s}_{i,n} &= \frac{(-1)^n}{2^n} \frac{1}{(i-n)! n!} \sum_{m=0}^{\infty} { n \choose m } \frac{(s!)^2}{(s-m)!(s+m-i)!}
\\ &= \frac{(-1)^n}{2^n} \frac{s^i}{(i-n)! n!} \sum_{m=0}^{\infty} { n \choose m } e^{- \frac{i^2-(2m+1)i+2m^2}{2s} + \CO(s^{-2})}
\\ &= \frac{(-1)^n s^i}{(i-n)! n!} \left( 1 - \frac{2i^2 -2(n+1)i + n(n+1)}{4s} + \CO(s^{-2}) \right)
\eld
\el
Note that in the $s \to \infty$ limit, $F^{s}_{i,n}$ scales as $\CO(s^i)$. This motivates us to introduce $\tilde{g}_i$ and $\tilde{F}^{s}_{i,n}$ as finite $s \to \infty$ quantity:
\bg
\tilde{g}_i = \frac{i!}{s^i} g_i \,, \quad \tilde{F}^{s}_{i,n} = \frac{i!}{s^i} F^{s}_{i,n} \label{eq:tgtFdef}
\eg
This particular scaling allows a simple expression for $\tilde{F}^{s}_{i,n}$ in the asymptotic limit $s \to \infty$.
\bl
\tilde{F}_{i,n} \equiv \lim_{s \to \infty} \tilde{F}^{s}_{i,n} = { i \choose n } (-1)^n \label{eq:tFmatDef}
\el
As a matrix, $\tilde{F}_{i,n}$ is a lower triangular infinite matrix which squares to the identity, i.e. $\sum_{n=0}^{\infty} \tilde{F}_{i,n} \tilde{F}_{n,j} = \delta_{i,j}$. Therefore, in the asymptotic limit $s \to \infty$
\eq
\tilde{g}_i = \sum_{n=0}^{\infty} \tilde{F}_{i,n} C_{\text{S}^n}\quad\rightarrow\quad C_{\text{S}^n}= \sum_{i=0}^{\infty} \tilde{F}_{n,i} \tilde{g}_i
\eqe
Inserting $\tilde{g}_0 = 1$ and $\tilde{g}_{i>0} = 0$ into the above equation indeed yields $C_{\text{S}^n} = 1$, which is the leading result in $\frac{1}{s}$. The subleading $\frac{1}{s}$ terms of $\tilde{F}^s_{i,n}$ is;
\bl
\bld
\tilde{F}^s_{i,n} 
&= \tilde{F}_{i,n} - \frac{2i^2 -2(n+1)i + n(n+1)}{4s} \tilde{F}_{i,n} + \CO(s^{-2}) \,.
\eld
\el
The inverse matrix up to the same asymptotic order can be computed using the formal matrix identity $(\iden - h)^{-1} = \sum_{i=0}^{\infty} h^i$.
\bl
\bld
\left( \tilde{F}^s \right)^{-1}_{n,i} &= \tilde{F}_{n,i} + \frac{1}{s} \sum_{j,k} \frac{2j^2 -2(k+1)j + k(k+1)}{4} \tilde{F}_{n,j} \tilde{F}_{j,k} \tilde{F}_{k,i} + \CO(s^{-2}) \,.
\\ &= \tilde{F}_{n,i} + \frac{1}{s} \frac{(-1)^i n!}{i!} \sum_{j,k} (-1)^{j+k} \frac{2j^2 -2(k+1)j + k(k+1)}{4 (n-j)! (j-k)! (k-i)!} + \CO(s^{-2}) \,.
\eld
\el
Therefore, the Wilson coefficients for minimal coupling up to this order is
\bl
\bld
C^{Min,s}_{\text{S}^{n}} = \left( \tilde{F}^{s} \right)^{-1}_{n,0} &= 1 + \frac{n!}{s} \sum_{j,k} (-1)^{j+k} \frac{2j^2 -2(k+1)j + k(k+1)}{4 (n-j)! (j-k)! k!} + \CO(s^{-2})
\\ &= 1 + \frac{n (n-1)}{4s} + \CO(s^{-2}) \,.
\eld
\el
It is also possible to work out higher order corrections to arbitrary order analytically using the procedures outlined above. For brevity, we only report the result for $C^{Min,s}_{\text{S}^{n}}$ up to $\CO(s^{-2})$ order obtained analytically;
\bl
C^{Min,s}_{\text{S}^{n}} &= 1 + \frac{n (n-1)}{4s} + \frac{(n^2-5n+10)n(n-1)}{32s^2} + \CO(s^{-3}) \,.
\el

\section{Properties of boosts}
\subsection{Approximating powers of boost generators $\vec{K}$} \label{app:boost2rot}
The Lorentz generators and their algebras are;
\begin{gather}
J^i = \left\{ \begin{aligned}
\frac{1}{2} \left( \s^i \right)_{\a}^{~\b} &\quad \text{Chiral}
\\ \frac{1}{2} \left( \s^i \right)^{\dot\a}_{~\dot\b} &\quad \text{Anti-chiral}
\end{aligned} \right. \, , \quad K^i = \left\{ \begin{aligned}
\frac{i}{2} \left( \s^i \right)_{\a}^{~\b} &\quad \text{Chiral}
\\ - \frac{i}{2} \left( \s^i \right)^{\dot\a}_{~\dot\b} &\quad \text{Anti-chiral}
\end{aligned} \right.
\\ [J^i, J^j] = i \e^{ijk} J^k \,,\quad [K^i, K^j] = - i \e^{ijk} J^k \,,\quad [J^i,K^j] = i \e^{ijk} K^k \,,
\end{gather}
where $\s^i$ are the Pauli matrices, $J^i = \half \e^{ijk} J^{jk}$ are the rotation generators, and $K^i = J^{i0}$ are the boost generators\footnote{Treatment of boost generators as rotation generators in general spacetime dimensions has been given in~\cite{Bautista:2019tdr}.}. The explicit form for the Lorentz group generators in the representation $(\frac{s}{2},\frac{s}{2})$ are obtained as a tensor sum of above.
\bl
\bld
J^i &= \half (\s^i \otimes \overbrace{\cdots \otimes \iden}^{2s-1} + \cdots + \overbrace{\iden \otimes \cdots}^{2s-1} \otimes \s^i)
\\ K^i &= \frac{i}{2} (\s^i \otimes \overbrace{\cdots \otimes \iden}^{2s-1} + \cdots + \overbrace{\iden \otimes \cdots}^{s-1} \otimes \s^i \otimes \overbrace{\cdots \otimes \iden}^{s}
\\ &\phantom{=asdfasdfasdfasdf} - \overbrace{\iden \otimes \cdots}^{s} \otimes \s^i \otimes \overbrace{\cdots \otimes \iden}^{s-1} - \cdots - \overbrace{\iden \otimes \cdots}^{2s-1} \otimes \s^i) \,.
\eld
\el
Since little group indices will always be symmetrised, the expressions for generators and their products can be simplified further. For this purpose, let us first fix the normalisations of the spinors for particles of unit mass at rest, where arrows $\uparrow$ and $\downarrow$ are the little group indices.
\bl
\bld
\ket{\bf{0}^\uparrow}_\a = \sket{\bf{0}^\uparrow}^{\dot\a} = {1 \choose 0} &,\quad \ket{\bf{0}^\downarrow}_\a = \sket{\bf{0}^\downarrow}^{\dot\a} = {0 \choose 1}
\\ \bra{\bf{0}^\uparrow}^\a = - \sbra{\bf{0}^\uparrow}_{\dot\a} = - (0 \quad 1) &,\quad \bra{\bf{0}^\downarrow}^\a = - \sbra{\bf{0}^\downarrow}_{\dot\a} = (1 \quad 0)
\eld
\el
The second line follows from the first line by adopting the definition $\e^{\uparrow \downarrow} = +1$. Adopting this normalisation, the generators and their products are simplified as below where $\stackrel{\cdot}{=}$ denotes numerical equivalence when inserted between bra and ket vectors of spin-$s$ states in the rest frame.
\bl
\bld
J^i &\stackrel{\cdot}{=} 2s \times \half \s^i \otimes \overbrace{\cdots \otimes \iden}^{2s-1}
\\ J^i J^j &\stackrel{\cdot}{=} 2s \times \frac{1}{2^2} \s^i \s^j \otimes \overbrace{\cdots \otimes \iden}^{2s-1} + (2s)(2s-1) \times \frac{1}{2^2} \s^i \otimes \s^j \otimes \overbrace{\cdots \otimes \iden}^{2s-2}
\\ K^i &\stackrel{\cdot}{=} 0
\\ K^i K^j &\stackrel{\cdot}{=} - 2s \times \frac{1}{2^2} \s^i \s^j \otimes \overbrace{\cdots \otimes \iden}^{2s-1} - \left[ 2 s (s-1) - 2 s^2 \right] \times \frac{1}{2^2} \s^i \otimes \s^j \otimes \overbrace{\cdots \otimes \iden}^{2s-2}
\eld
\el
The symmetrisation argument can be used to show that $(K)^{2n+1} (J)^m \stackrel{\cdot}{=} 0$, so only even powers of $\vec{K}$ need to be worked out. The contribution with largest $s$ dependence will be the contribution where all Pauli matrices are allotted to different spinor indices, given that $s>n$. The coefficient for such a contribution can be worked out from simple combinatorics.
\bl
\overbrace{J \cdots J}^{2n} &\stackrel{\cdot}{\simeq} \frac{(2s)!}{(2s-2n)!} \frac{1}{2^{2n}} \overbrace{\s \otimes \cdots}^{2n} \otimes \overbrace{\iden \otimes \cdots \otimes \iden}^{2s-2n} + \cdots
\\ \overbrace{K \cdots K}^{2n} &\stackrel{\cdot}{\simeq} \left[ \sum_{m=0}^{2n} (-1)^m {2n \choose m} \frac{(s!)^2}{(s-m)! (s-2n+m)!} \right] \frac{(-1)^n}{2^{2n}} \overbrace{\s \otimes \cdots}^{2n} \otimes \overbrace{\iden \otimes \cdots \otimes \iden}^{2s-2n} + \cdots \nn
\\ &= (2n)! {s \choose n} \frac{1}{2^{2n}} \overbrace{\s \otimes \cdots}^{2n} \otimes \overbrace{\iden \otimes \cdots \otimes \iden}^{2s-2n} + \cdots
\el
Comparing the two yields the relation
\bl
\frac{1}{(2n)!} \left( \vec{\l} \cdot \vec{K} \right)^{2n} &\stackrel{\cdot}{=} \frac{(2s-2n)!}{(2s)!} {s \choose n} \left( \vec{\l} \cdot \vec{J} \right)^{2n} + ( \vec{\l} )^2 F_{2n-2} (\vec{\l} \cdot \vec{J}) \label{eq:Boost2Rot}
\el
where $F_{2m}(x)$ is some even polynomial of degree $2m$. The appearance of the factor $(\vec{\l})^2$ follows from anti-commutator of Pauli matrices; $\s^i \s^j + \s^j \s^i = 2 \delta^{ij}$.

\section{Universality at one-loop} \label{sec:2PMUniv}

\begin{figure}
\centering
\includegraphics[width=0.4\textwidth]{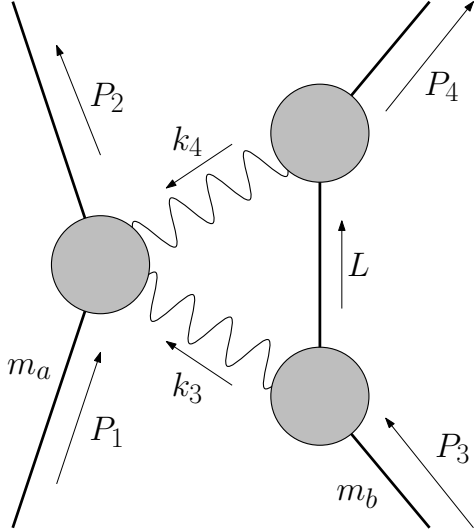} \caption{The triangle cut used to compute the 2 PM potential.} \label{fig:guevaratriangle}
\end{figure}

In this section, we will only be interested in the  $ G^2 \hbar^0 |\vec{q}|^{-1}$ effects which can be cleanly captured by the $t$-channel triangle in the HCL limit. To compute the integral coefficient, we apply the unitarity cut approach \cite{Bern:1994cg,Bern:1994zx} especially by Forde~\cite{Forde:2007mi}, where the contributions of each integral is separated by their distinct set of propagators. By putting these propagators on-shell unitarity dictates that the result must be given by the product of tree-amplitudes. In our case, the triangle cut, we have the product of two minimal coupling three-point amplitude and a gravitational Compton amplitude, as illustrated in fig.\ref{fig:guevaratriangle}. The triangle coefficient can then be captured by removing the contributions from the box integrals. 

Compared to tree level, the new feature at one-loop is the gravitational Compton amplitude, involving two massive spinning particles and two massless gravitons:
$$\includegraphics[scale=0.6]{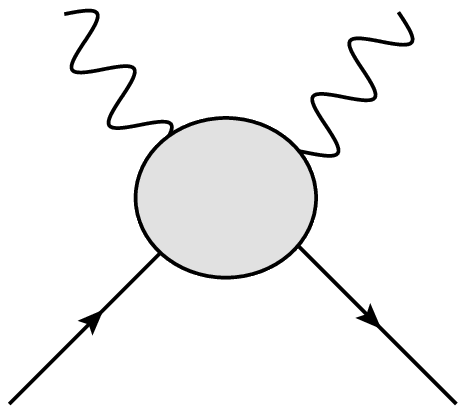}$$
For Kerr black holes, we will be interested in the Compton amplitude which correspond to the four-point extension of the three-point minimal coupling. More precisely, the residue of the massive pole must yield the product of three-point minimal coupling discussed previously. However as shown in \cite{Arkani-Hamed:2017jhn} and \cite{Chung:2018kqs} for $s>2$,  due to polynomial ambiguities, factorization constraints do not uniquely determine the gravitational Compton amplitude. More precisely, by matching to the factorization pole in all three channels, for $s\leq2$ we can find a solution:
\begin{equation}\label{eq:Lower_spin_Compton}
M_4^{(s \leq 2)}\left( \BS{1}^s, -\BS{2}^s, k_3^{-2}, k_4^{+2}\right) = - \frac{\MixLeft{k_3}{p_1}{k_4}^{4}}{\MixLeft{k_3}{p_1}{k_3} \MixLeft{k_4}{p_1}{k_4} \MixLeft{k_4}{k_3}{k_4}} \left( \frac{-\AB{\BS{1}k_3}\SB{\BS{2} k_4} + \AB{\BS{2}k_3}\SB{ \BS{1} k_4}}{\MixLeft{k_3}{p_1}{k_4}} \right)^{2s}
\end{equation}
where $|k_4\rangle, |k_4]$ and $|k_3\rangle, |k_3]$ are massless spinors for the massless propagators. Importantly, the result does not contain any $\frac{1}{m}$ factors. This has two important implications: 1. One can take $m\rightarrow0$ limit smoothly, indicating that the spinning particle has a point-like description. 2. Since pure polynomial terms must have $\frac{1}{m^n}$ factors simply on dimensional grounds, they can be considered as finite size effects and do not mix with eq.(\ref{eq:Lower_spin_Compton}). For $s>2$, the situation is drastically different. The amplitude takes the form (see \cite{Chung:2018kqs} ):
\eq
M_4^{(s> 2)}\left( \BS{1}^s, -\BS{2}^s, k_3^{-2}, k_4^{+2}\right) = - \frac{\MixLeft{k_3}{p_1}{k_4}^{4}}{\MixLeft{k_3}{p_1}{k_3} \MixLeft{k_4}{p_1}{k_4} \MixLeft{k_4}{k_3}{k_4}} \left(\frac{\langle \mathbf{12}\rangle {-}[\mathbf{12}]}{m}{+}\cdots \right)^{2s}{+}\cdots\,,
\eqe
where we've only listed the leading term in propagators and $\frac{1}{m}$ expansion. We see that unlike $s\leq2$, here the leading piece already contains non-trivial $\frac{1}{m}$ dependence, thus making the separation of finite size effects from that of what is dictated by factorization operationally meaningless.

\subsection{Universality for $s\leq2$}
For $s \leq 2$ particles, we insert the Compton amplitude eq.\eqref{eq:Lower_spin_Compton} into the computation of the triangle integral in the HCL limit. 
Taking spin 1 and spin 2 as an example, the triangle integral in the HCL limit yields the following coefficients for the anti-chiral basis (defined in eq.(\ref{eq:1PMAnsatzDemo2})):\footnote{Here, $\epsilon = \sqrt{\rho^2-1}$}
\begin{subequations}
\begin{itemize}[leftmargin=*]
\item $S_a^0 S_b^0$
\begin{equation}\label{eq:curly_A00}
\begin{split}
\frac{\mathcal{A}_{0,0}^{s=1}}{m_a^2m_b^2} &= -\frac{24 \pi ^2 G^2  \left(m_a+m_b\right)}{|\vec{q}|} + O(\epsilon^2) \\
\frac{\mathcal{A}_{0,0}^{s=2}}{m_a^2m_b^2} &= -\frac{24 \pi ^2 G^2  \left(m_a+m_b\right)}{|\vec{q}|} + O(\epsilon^2)
\end{split}
\end{equation}
\item $S_a^1 S_b^0$
\begin{equation}
\begin{split}
\frac{\mathcal{A}_{1,0}^{s=1}}{m_a^2m_b^2} &= \frac{4 \pi ^2 G^2  \left(4 m_a+3 m_b\right)}{|\vec{q}| \epsilon }+\frac{24 \pi ^2 G^2  \left(m_a+m_b\right)}{|\vec{q}|} + O(\epsilon)\\
\frac{\mathcal{A}_{1,0}^{s=2}}{m_a^2m_b^2} &= \frac{8 \pi ^2 G^2  \left(4 m_a+3 m_b\right)}{|\vec{q}| \epsilon }+\frac{48 \pi ^2 G^2  \left(m_a+m_b\right)}{|\vec{q}|} + O(\epsilon)
\end{split}
\end{equation}
\item $S_a^1 S_b^1$
\begin{equation}
\begin{split}
\frac{\mathcal{A}_{1,1}^{s=1}}{m_a^2m_b^2} &= \frac{4 \pi ^2 G^2  \left(m_a+m_b\right)}{|\vec{q}| \epsilon ^2}+\frac{4 \pi ^2 G^2  \left(m_b-m_a\right)}{|\vec{q}| \epsilon }+\frac{14\pi ^2 G^2 \left(m_a+m_b\right)}{|\vec{q}|} + O(\epsilon) \\
\frac{\mathcal{A}_{1,1}^{s=2}}{m_a^2m_b^2} &= \frac{16 \pi ^2 G^2  \left(m_a+m_b\right)}{|\vec{q}| \epsilon ^2}+\frac{16 \pi ^2 G^2  \left(m_b-m_a\right)}{|\vec{q}| \epsilon }+\frac{56\pi ^2 G^2  \left(m_a+m_b\right)}{|\vec{q}|} + O(\epsilon)
\end{split}
\end{equation}
\item $S_a^2 S_b^0$ 
\begin{equation}\label{eq:curly_A20}
\begin{split}
\frac{\mathcal{A}_{2,0}^{s=1}}{m_a^2m_b^2} &= -\frac{\pi ^2 G^2 \left(m_a+m_b\right)}{|\vec{q}| \epsilon ^2}-\frac{2 \pi ^2 G^2 \left(4 m_a+3 m_b\right)}{|\vec{q}| \epsilon}-\frac{\pi ^2 G^2 \left(34 m_a+27 m_b\right)}{2 |\vec{q}|} + O(\epsilon) \\
\frac{\mathcal{A}_{2,0}^{s=2}}{m_a^2m_b^2} &= -\frac{6 \pi ^2 G^2 \left(m_a+m_b\right)}{|\vec{q}| \epsilon ^2}-\frac{12 \pi ^2 G^2  \left(4 m_a+3 m_b\right)}{|\vec{q}| \epsilon}-\frac{3 \pi ^2 G^2 \left(34 m_a+27 m_b\right)}{|\vec{q}|}+ O(\epsilon)
\end{split}
\end{equation}
\item $S_a^2 S_b^1$
\begin{equation}
\begin{split}
\frac{\mathcal{A}_{2,1}^{s=1}}{m_a^2m_b^2} &= -\frac{\pi ^2 G^2 \left(m_a+m_b\right)}{|\vec{q}| \epsilon ^2}-\frac{\pi ^2 G^2 \left(m_a+6 m_b\right)}{2 q \epsilon }-\frac{\pi^2 G^2 \left(4 m_a+11 m_b\right)}{2 |\vec{q}|} + O(\epsilon) \\
\frac{\mathcal{A}_{2,1}^{s=2}}{m_a^2m_b^2} &= -\frac{12 \pi ^2 G^2 \left(m_a+m_b\right)}{|\vec{q}| \epsilon ^2}-\frac{6 \pi ^2 G^2 \left(m_a+6 m_b\right)}{|\vec{q}| \epsilon }-\frac{6\pi ^2 G^2 \left(4 m_a+11 m_b\right)}{|\vec{q}|} + O(\epsilon)
\end{split}
\end{equation}
\item $S_a^2 S_b^2$
\begin{equation}\label{eq:curly_A22}
\begin{split}
\frac{\mathcal{A}_{2,2}^{s=1}}{m_a^2m_b^2} &= \frac{\pi ^2 G^2 \left(m_a+m_b\right)}{8 |\vec{q}| \epsilon ^2}+\frac{\pi ^2 G^2\left(m_b-m_a\right)}{4 |\vec{q}| \epsilon }+\frac{13\pi ^2 G^2 \left(m_a+m_b\right)}{32 |\vec{q}|} + O(\epsilon) \\
\frac{\mathcal{A}_{2,2}^{s=2}}{m_a^2m_b^2} &= \frac{9 \pi ^2 G^2 \left(m_a+m_b\right)}{2 |\vec{q}| \epsilon ^2}+\frac{9 \pi ^2 G^2 \left(m_b-m_a\right)}{|\vec{q}| \epsilon }+\frac{117\pi ^2 G^2 \left(m_a+m_b\right)}{8 |\vec{q}|} + O(\epsilon)
\end{split}
\end{equation}
\end{itemize}
\end{subequations}
Importantly, the $\mathcal{A}$ coefficients satisfy the following identity:
\begin{equation}
\mathcal{A}^{s=2}_{i \leq 2, j \leq 2} = \frac{4!^2}{(4-i)!(4-j)!} \frac{(2-i)!(2-j)!}{2!^2} \mathcal{A}^{s=1}_{i\leq 2,j\leq 2} \,.
\end{equation}
The factors in front of $\mathcal{A}^{s=1}_{i\leq 2,j\leq 2}$ are the combinatoric factors defined in eq.(\ref{eq:universality1}), of spin 2 devided by that of spin 1. Thus we again find that at one-loop, the classical contribution extracted from minimally coupled spin 1 and 2 respects universality as defined in eq.(\ref{eq:universality1}).  Although we've only displayed up to $O(\epsilon)^0$ for brevity in eq.\eqref{eq:curly_A00} to eq.\eqref{eq:curly_A22}, the universality behaviour is satisfied to all orders in $\epsilon$. Similarly, we've also checked that universality also holds for spin-$1/2$ and $3/2$.

\subsection{Universality for $s>2$}
Given that universality is found to hold for minimal coupling with $s\leq2$, we naturally ask whether there exists some higher spin extension of minimally coupled gravitational Compton amplitude, such that universality is respected. A straightforward BCFW construction of the Compton amplitude yields a non-local expression for $s>2$ (see also~\cite{Johansson:2019dnu}). However since we are interested in only the classical part of the triangle integral, a priori it is not clear whether this would yield a problematic potential. 
On the other hand, a manifestly local, albeit non-unique, higher-spin extension was given in eq.(5.24) of \cite{Chung:2018kqs}. 
We analyse both cases separately in this section and show that universality can be maintained.

\subsubsection{The BCFW Compton amplitude} \label{sec:BCFWUniv}
First, we start from the BCFW representation of the Compton amplitude, which is constructed from the $\langle k_3, k_4]$ shift:
\begin{equation}\label{eq:BCFW_rep}
\begin{split}
&M^{\text{BCFW}}_4(\BS{1}^s, \BS{2}^s, k_3^{-2}, k_4^{+2})\\
=& \frac{ \hat{M}_3(\BS{1}^s, -\BS{\hat{P}}_{14}^{s}, \hat{k}_4^{+2})\hat{M}_3(\BS{2}^s, \BS{\hat{P}}_{14}^{s}, \hat{k}_3^{-2}) }{\MixLeft{k_4}{p_1}{k_4}} + \frac{ \hat{M}_3(\BS{1}^s, -\BS{\hat{P}}_{13}^{s}, \hat{k}_3^{-2})\hat{M}_3(\BS{2}^s, \BS{\hat{P}}_{13}^{s}, \hat{k}_4^{+2}) }{\MixLeft{k_3}{p_1}{k_3}}\\
=& - \frac{\MixLeft{k_3}{p_1}{k_4}^{4}}{\MixLeft{k_3}{p_1}{k_3} \MixLeft{k_4}{p_1}{k_4} \MixLeft{k_4}{k_3}{k_4}} \left( \frac{\AB{\BS{1}k_3}\SB{\BS{2} k_4} + \AB{\BS{2}k_3}\SB{ \BS{1} k_4}}{\MixLeft{k_3}{p_1}{k_4}} \right)^{2s}
\end{split}
\end{equation}
Taking $p_2 \rightarrow -p_2$, it indeed recovers \eqref{eq:Lower_spin_Compton}. Note that while for $s>2$ the expression is non-local, it factorizes correctly on all three channels. 
Since it is just an extrapolation of eq.\eqref{eq:Lower_spin_Compton} to $s>2$, not surprisingly the HCL limit of the triangle integral satisfy universality for all spins. Take $\mathcal{A}_{4,0}$ as an example:
\begin{equation}
\begin{split}
\frac{\mathcal{A}^{s=3}_{4,0}}{m_a^2 m_b^2} &= -\frac{225 \pi ^2 G^2 \left(m_a+m_b\right)}{8 |\vec{q}| \epsilon ^2}-\frac{75 \pi ^2 G^2 \left(4 m_a+3 m_b\right)}{2 |\vec{q}| \epsilon }-\frac{225 \pi ^2 G^2 \left(23 m_a+16 m_b\right)}{16 |\vec{q}|} + O(\epsilon)\\
\frac{\mathcal{A}^{s=2}_{4,0}}{m_a^2 m_b^2} &= \frac{15 \pi ^2 G^2 \left(m_a+m_b\right)}{8 |\vec{q}| \epsilon ^2}-\frac{5 \pi ^2 G^2 \left(4 m_a+3 m_b\right)}{2 |\vec{q}| \epsilon }-\frac{15 \pi ^2 G^2 \left(23 m_a+16 m_b\right)}{16 |\vec{q}|} + O(\epsilon)
\end{split}
\end{equation}
We find that $\mathcal{A}^{s=3}_{4,0} = 15 \mathcal{A}^{s=2}_{4,0} = \frac{6!^2}{(6-4)!(6-0)!} \frac{(4-4)!(4-0)!}{4!^2} \mathcal{A}_{4,0}$, which indeed satisfy the universality relation.

An interesting self-consistency test is the following; assuming universality also holds at one-loop, then the computations done for $s\leq2$ would be sufficient to capture the correct potential involving operators $S^{n \leq 4}$ by RHS of eq.\eqref{eq:BcoeffSol1}. Here, we take the BCFW Compton amplitude in the polarization tensor basis and take the classical-spin limit of the triangle integral. In the basis of eq.(\ref{eq:SpinInSpinorVar}) we find:
\begin{itemize}[leftmargin=*]
\item $S_a^0 S_b^0$
\begin{equation}
B^{\text{BCFW}}_{0,0} = \frac{24  \pi^2 G^2 m_a^2 m_b^2 \left(m_a+m_b\right)}{|\vec{q}|} + O(\epsilon)^2
\end{equation}

\item $S_a^1 S_b^0$
\begin{equation}
B^{\text{BCFW}}_{1,0} = -\frac{4 \pi ^2 G^2 m_a m_b^2 \left(4 m_a+3 m_b\right)}{|\vec{q}| \epsilon } - \frac{12 \pi ^2 G^2  m_a m_b^2 \left(4 m_a+3 m_b\right)}{|\vec{q}|} \epsilon + O(\epsilon)^3
\end{equation}
\item $S_a^2 S_b^0$
\begin{equation}
\begin{split}
B^{\text{BCFW}}_{2,0} 
&=\frac{2 \pi ^2 G^2 m_b^2 \left(m_a+m_b\right)}{|\vec{q}| \epsilon^2}+\frac{\pi ^2 G^2 m_b^2 \left[ \left(22+\frac{12}{2 s-1}\right) m_a+ \left(15+\frac{12}{2 s-1} \right) m_b\right] }{|\vec{q}|} + O(\epsilon)^2\\
& = \frac{2 \pi ^2 G^2 m_b^2 \left(m_a+m_b\right)}{|\vec{q}| \epsilon^2}+\frac{\pi ^2 G^2 m_b^2 \left( 22 m_a+ 15 m_b\right) }{|\vec{q}|}+ O\left(\frac{1}{s}\right) + O(\epsilon)^2
\end{split}
\end{equation}
\item $S_a^3 S_b^0$
\begin{equation}
\begin{split}
&B^{\text{BCFW}}_{3,0} \\
 =& -\frac{ \pi ^2 G^2 m_b^2 \left( 1+\frac{2}{2 s-1} \right) \left(4 m_a+3 m_b\right)}{ |\vec{q}| \epsilon m_a}-\frac{\pi ^2 G^2  m_b^2 \left[\left( 11+\frac{24}{2 s-1} \right) m_a+ \left(\frac{13}{2}+\frac{18}{2 s-1}\right) m_b\right]}{ |\vec{q}| m_a} \epsilon + O(\epsilon)^3\\
 =& - \frac{ \pi ^2 G^2 m_b^2 \left(4 m_a+3 m_b\right)}{ |\vec{q}| m_a \epsilon}-\frac{\pi ^2 G^2  m_b^2 \left(22 m_a+13 m_b\right)}{2 |\vec{q}| m_a} \epsilon + O\left(\frac{1}{s}\right) + O(\epsilon)^3
\end{split}
\end{equation}
\item $S_a^4 S_b^0$
\begin{equation}
\begin{split}
& B^{\text{BCFW}}_{4,0}  \\
=& \frac{\pi ^2 G^2 \left(\frac{1}{4} + \frac{1}{2 s-1}\right) m_b^2 \left(m_a+m_b\right)}{|\vec{q}| m_a^2 \epsilon^2 } + \frac{ \pi^2 G^2 m_b^2\left[\frac{152 s^3-28 s^2-390 s+87}{8 (1-2 s)^2 (2 s-3)}m_a + \frac{3 \left(4 s^3-11 s+1\right)}{(1-2 s)^2 (2 s-3)}m_b\right] }{ |\vec{q}|m_a^2} + O(\epsilon)^2\\
=& \frac{\pi ^2 G^2 m_b^2 \left(m_a+m_b\right)}{4 |\vec{q}| m_a^2 \epsilon^2 } + \frac{ \pi^2 G^2 m_b^2\left( 19 m_a + 12 m_b\right) }{ 8m_a^2  |\vec{q}|} + O\left(\frac{1}{s}\right) + O(\epsilon)^2
\end{split}
\end{equation}
\end{itemize}
We find that it indeed reproduces the RHS of eq.\eqref{eq:BcoeffSol1}.
We also comment that universality of BCFW representation is based on the same combinatoric structure at tree level analysed in section \ref{sec:1PMrevisit}.

\subsubsection{The $s>2$ local Compton amplitude} \label{sec:localUniv}
Now, we turn to the local representation of the Compton amplitude given in eq.(5.24) of \cite{Chung:2018kqs}. We first ask if universality is respected when comparing with the known Compton amplitudes ($s\leq2$) for coefficients $\mathcal{A}_{0,0}$ to $\mathcal{A}_{4,4}$, and if not can the situation be rectified by inclusion of suitable polynomial terms. 

We analyze the coefficients $\mathcal{A}_{i,j}$ extracted from the gravitational Compton amplitude for $s=6$ in \cite{Chung:2018kqs}. For $i,j\leq4$, we expect to match with that computed from $s\leq2$ via universality. In the following, we give a few examples.
\begin{enumerate}[leftmargin=*]
\item $i,j<4$ $(S_a^{i <4}S_b^{j <4})$: 
 
\begin{equation}
\frac{\mathcal{A}_{2,0}^{\text{local}, s=6}}{m_a^2 m_b^2} = -\frac{66 \pi ^2 G^2 \left(m_a+m_b\right)}{|\vec{q}| \epsilon ^2}-\frac{132 \pi ^2 G^2 \left(4 m_a+3 m_b\right)}{|\vec{q}| \epsilon}-\frac{33 \pi ^2 G^2 \left(34 m_a+27 m_b\right)}{|\vec{q}|} + O(\epsilon)
\end{equation}
Comparing with eq.\eqref{eq:curly_A20}, one finds
\begin{equation}
\mathcal{A}_{2,0}^{\text{local},s=6} = \frac{12!}{(12-2)!}\frac{(4-2)!}{4!} \mathcal{A}_{2,0}^{s=2} =  \frac{12!}{(12-2)!}\frac{(2-2)!}{2!} \mathcal{A}_{2,0}^{s=1}
\end{equation}
where universality between the local amplitude and eq.\eqref{eq:Lower_spin_Compton} is satisfied. Similarly, comparing between different higher spins, we also find universality. For example for $s=4,6$  
\begin{equation}
\begin{split}
\frac{\mathcal{A}_{2,0}^{\text{local}, s=4}}{m_a^2 m_b^2} 
&= - \frac{28 \pi^2  G^2 \left(m_a+m_b\right)}{ |\vec{q}| \epsilon ^2} - \frac{56 \pi^2  G^2 \left(4 m_a+3 m_b\right)}{|\vec{q}| \epsilon }-\frac{14 \pi^2  G^2 \left(34 m_a+27 m_b\right)}{ |\vec{q}|}  + O(\epsilon) \\
&= \frac{8!}{(8-2)!}\frac{(12-2)!}{12!}\frac{\mathcal{A}_{2,0}^{\text{local}, s=6}}{m_a^2 m_b^2} 
\end{split}
\end{equation}
\item $i=4,j<4$ $(S_a^{i =4}S_b^{j <4})$ effects: 

Operators with $S^4$ are the highest degree for which spin-2 particles can probe. Here, unlike the BCFW higher-spin extension, we indeed find discrepancy with that from universality. Take for example $\mathcal{A}_{4,0}$: \begin{equation}
\begin{split}
\frac{\mathcal{A}_{4,0}^{\text{local},s=6} }{m_a^2 m_b^2} 
&= -\frac{7425 \pi ^2 G^2 \left(m_a+m_b\right)}{8 |\vec{q}| \epsilon ^2}-\frac{2475 \pi ^2 G^2 \left(4 m_a+3 m_b\right)}{2 |\vec{q}| \epsilon }-\frac{1485 \pi ^2 G^2 \left(460 m_a+317 m_b\right)}{64 |\vec{q}|} \\
& \qquad \qquad+ O(\epsilon) \\
&= \left\lbrace 495-\frac{297 \epsilon ^2 m_b}{8 \left(m_a+m_b\right)} + O(\epsilon^3) \right\rbrace \frac{\mathcal{A}_{4,0}^{s=2} }{m_a^2 m_b^2} 
\end{split}
\end{equation}
We see that the $s=6$ and $s=2$ results no longer differ by an overall combinatoric factor. In other words, universality is lost. However, as discussed in \cite{Chung:2018kqs}, given the polynomial ambiguity of Compton amplitudes beyond $s>2$, we can restore universality simply by adding local contact terms. Indeed by adding 
\eq
M_{\text{contact}}^{s=6} = -\frac{3}{2} (8\pi G) m^2 \binom{12}{4} \left(\frac{\AB{k_3 \BS{2}}\SB{k_4 \BS{1}} + \AB{k_3 \BS{1}}\SB{k_4 \BS{2}}}{2m^2}\right)^4\left(\frac{\AB{\BS{12}} - \SB{\BS{12}}}{2m}\right)^{8}
\eqe
to the spin-6 Compton amplitude, we find:
\begin{equation}
\begin{split}
&\frac{\mathcal{A}_{4,0}^{\text{local+contact}, s=6}}{m_a^2 m_b^2} \\
= &-\frac{7425 \pi ^2 G^2 \left(m_a+m_b\right)}{8 |\vec{q}| \epsilon ^2}-\frac{2475 \pi ^2 G^2 \left(4 m_a+3 m_b\right)}{2 |\vec{q}| \epsilon }-\frac{7425 \pi ^2 G^2 \left(23 m_a+16 m_b\right)}{16 |\vec{q}|} + O(\epsilon)\\
= &\frac{12!}{(12-4)!} \frac{(4-4)!}{4!} \times \frac{\mathcal{A}_{4,0}^{s=2} }{m_a^2 m_b^2} 
\end{split}
\end{equation}
We now see that a correct contact term makes $A_{i=4,j<4} = A^{\text{local + contact}}_{i=4,j<4}$. For generic spins, the suitable contact term is:
\begin{equation}\label{eq:General_s_contact}
M_{\text{contact}}^{s} = -\frac{3}{2} (8\pi G) m^2 \binom{2s}{4} \left(\frac{\AB{k_3 \BS{2}}\SB{k_4 \BS{1}} + \AB{k_3 \BS{1}}\SB{k_4 \BS{2}}}{2m^2}\right)^4\left(\frac{\AB{\BS{12}} - \SB{\BS{12}}}{2m}\right)^{2s-4}
\end{equation}
Note that this contact term is constructed in a way that it only modifies the behaviour of $S_a^{i \geq 4}S_b^{j \geq 4}$ and we do not need to worry about breakdown of universality for $S_a^{i =4}S_b^{j <4}$. 
\item $i>4,j>4$ $(S_a^{i > 4}S_b^{j > 4})$: 

Having patched up our local expression such that universality is respected for $\mathcal{A}_{i,j}$ with $i,j\leq4$, we now turn to its fate for $i,j>4$. Again, focusing on spin-6, we find
\begin{equation}
\begin{split}
&\frac{\mathcal{A}_{5,0}^{\text{local+contact}, s=6}}{m_a^2 m_b^2} \\
=&\frac{3465 \pi ^2 G^2 \left(m_a+m_b\right)}{2 |\vec{q}| \epsilon ^2}+\frac{17325 \pi ^2 G^2 \left(4 m_a+3 m_b\right)}{8 |\vec{q}| \epsilon}+\frac{693 \pi ^2 G^2 \left(107 m_a+72 m_b\right)}{4 |\vec{q}|}  + O(\epsilon)\\
&\frac{\mathcal{A}_{5,0}^{\text{local+contact}, s=3}}{m_a^2 m_b^2} \\
=&\frac{105 \pi ^2 G^2 \left(m_a+m_b\right)}{8 |\vec{q}| \epsilon ^2}+\frac{525 \pi ^2 G^2 \left(4 m_a+3 m_b\right)}{32 |\vec{q}| \epsilon}+\frac{21 \pi ^2 G^2 \left(107 m_a+72 m_b\right)}{16 |\vec{q}|} + O(\epsilon) \\
=& \frac{6!}{(6-5)!} \frac{(12-5)!}{12!} \frac{\mathcal{A}_{5,0}^{\text{local+contact}, s=6}}{m_a^2 m_b^2}
\end{split}
\end{equation}
where universality is still preserved. Here, we conclude that even with the modification from the contact terms eq.\eqref{eq:General_s_contact} universality is still preserved by all spins.
\end{enumerate}

%

\section{Residue integral representation for three-point amplitude} \label{sec:resrep}
To prove that minimal coupling reproduces Wilson coefficients of Kerr black holes after Hilbert-space matching, we first note that chiral spinor brackets and anti-chiral spinor brackets can be exchanged when momentum of in-state $p_1$ and out-state $p_1'$ are the same, $p_1 = p_1'$.
\bl
\la \mathbf{1'} \bf{1} \ra = - [\mathbf{1'1}]
\el
Following the kinematical set-up in section~\ref{sec:MinCoupBH}, we go to the frame where $p_1 = p_1'$ is at rest. The Hilbert space matching of minimal coupling becomes
\bl
\bld
\la \bf{21} \ra^{2s} &= \bra{\mathbf{1'}}^{2s} e^{i \frac{q \cdot K}{m}} \ket{\bf{1}}^{2s} = \bra{\mathbf{1'}}^{2s} e^{- \frac{q \cdot S}{m}} \ket{\bf{1}}^{2s} \simeq \ve^\ast ({\mathbf{1'}}) e^{- \frac{q \cdot S}{m}} \ve({\bf{1}})
\\ [ \bf{21} ]^{2s} &= \sbra{\mathbf{1'}}^{2s} e^{i \frac{q \cdot K}{m}} \sket{\bf{1}}^{2s} = \sbra{\mathbf{1'}}^{2s} e^{\frac{q \cdot S}{m}} \sket{\bf{1}}^{2s} \simeq \ve^\ast ({\mathbf{1'}}) e^{ \frac{q \cdot S}{m}} \ve({\bf{1}})
\eld \label{eq:MinKerr}
\el
which is the chiral(anti-chiral) basis version of eq.\eqc{eq:3ptBoost2Rot}. The relations $K = i J = i S$ for chiral spinors and $K = - i J = - i S$ for anti-chiral spinors has been used, and $\simeq$ in the above expression denotes equivalence up to normalisation. In other words, \emph{minimal coupling corresponds to unity Wilson coefficients} $C_{\text{S}^n_{eff}} = 1$. The same conclusion has been reached from heavy particle effective theory (HPET) point of view in~\cite{Aoude:2020onz}.

The map can be generalised to arbitrary $C_{\text{S}_{eff}^n}$. Using the expression for $\frac{q \cdot S}{m}$ in the chiral basis~\cite{Chung:2018kqs},
\bl
\bld
g_i \la \bf{21} \ra^{2s-i} \left( \frac{x \la \bf{2}q \ra \la q \bf{1} \ra}{m} \right)^i &= \frac{2^i (2s-i)! g_i}{(2s)!} \bra{\bf{2}}^{2s} \left( \frac{q \cdot S}{m} \right)^i \ket{\bf{1}}^{2s}
\\ &= \frac{2^i \hat{g}_i}{i!} \bra{\bf{2}}^{2s} \left( \frac{q \cdot S}{m} \right)^i \ket{\bf{1}}^{2s} \,,
\eld
\el
where we have introduced the notation $g_i = {2s \choose i} \hat{g}_i$. The above expression can be matched to eq.\eqc{eq:EFT3ptWCeff}, which expresses the amplitude through $C_{\text{S}_{eff}^n}$.
\bl
\bld
\frac{2^i \hat{g}_i}{i!} \bra{\bf{2}}^{2s} \left( \frac{q \cdot S}{m} \right)^i \ket{\bf{1}}^{2s} &= \frac{2^i \hat{g}_i}{i!} \bra{\mathbf{1'}}^{2s} e^{- \frac{q \cdot S}{m}} \left( \frac{q \cdot S}{m} \right)^i \ket{\bf{1}}^{2s}
\\ &= \bra{\mathbf{1'}}^{2s} \sum_n \frac{C_{\text{S}_{eff}^n}}{n!} \left( - \frac{q \cdot S}{m} \right)^n \ket{\bf{1}}^{2s} \,,
\eld
\el
yielding the relation
\bl
g_i &= {2s \choose i} \hat{g}_i = \frac{1}{2^i} {2s \choose i} \sum_{n} (-1)^n {i \choose n} C_{\text{S}_{eff}^n} \,. \label{eq:WCeff2gi}
\el


We may ask if there is an expression for the three-point amplitude that directly expresses the ampitude in terms of $C_{\text{S}_{eff}^n}$. Such an alternative expression 
can be found by adopting following definitions.
\bl
\bld
\bar{u}(p_2) u(p_1) &= \frac{ [ \bf{21} ] - \la \bf{21} \ra }{2m}
\\ S_{1/2}^\m &= \half \bar{u}(p_2) \g^\m \g_5 u(p_1) = - \frac{1}{4m} \left( \sbra{\bf{2}} \bar{\s}^\m \ket{\bf{1}} + \bra{\bf{2}} \s^\m \sket{\bf{1}} \right)
\eld
\el
The definition for spin vector $S^\m_{1/2}$, which can be considered as the scaled spin vector $\frac{S_s^\m}{2s}$ of the full spin vector of a spin-$s$ particle $S_s^\m$, has been adopted from Holstein and Ross~\cite{Holstein:2008sx} with a sign choice that matches to our conventions. An extra factor of $\frac{1}{2m}$ has been inserted as a normalisation condition $\bar{u}_I (p) u^J (p) = \delta_I^J$. Setting the momentum conservation condition as $p_2 = p_1 + q$, we propose the following \emph{residue integral representation} of three-point amplitude which expresses spin-$s$ amplitude as $2s$ power of spin-$\half$ amplitude.
\bl
\bld
M_s^{2 \eta} &= \frac{\k m x^{2\eta}}{2} \oint \frac{dz}{2\pi i z} \left( \sum_{n=0}^\infty C_{\text{S}_r^n} z^n \right) \left( \bar{u}(p_2) u(p_1) - \eta \frac{q \cdot S_{1/2}}{m z} \right)^{2s}
\\ &= \frac{\k m x^{2\eta}}{2} \oint \frac{dz}{2\pi i z} \left( \sum_{n=0}^\infty C_{\text{S}_r^n} z^n \right) \left( \frac{ [ \bf{21} ] - \la \bf{21} \ra }{2m} + \frac{\eta}{z} \frac{ \sbra{\bf{2}} q \ket{\bf{1}} + \bra{\bf{2}} q \sket{\bf{1}} }{4m^2} \right)^{2s}\,.
\eld \label{eq:EFT3ptNew}
\el
Here the contour encircles the origin, and the contour integral merely serves the auxiliary function of extracting the right combinatoric factors. 
For positive helicity $\eta = +1$, this expression becomes
\bl
M_s^{+2} &= (-1)^{2s} \frac{\k m x^2}{2} \oint \frac{dz}{2\pi i z} \left( \sum_{n=0}^\infty C_{\text{S}_r^n} z^n \right) \left( \frac{ \la \bf{21} \ra }{m} + \frac{z-1}{z} \frac{x \la \bf{2}3 \ra \la 3 \bf{1} \ra}{2m^2} \right)^{2s} \,.
\el
Using the binomial expansion leaves the following residue integral to be worked out.
\bl
\oint \frac{dz}{2\pi i z} z^n \left( \frac{z-1}{z} \right)^i = (-1)^i \sum_{j=0}^{i} \oint \frac{dz}{2\pi i z} { i \choose j } (-z)^j z^{n-i} = (-1)^n { i \choose n } \,.
\el
One then finds:
\bl
\bld
M_s^{+2} &= (-1)^{2s} \frac{\k m x^2}{2} \sum_{i=0}^{2s} { 2s \choose i} \left[ \frac{1}{2^i} \sum_{n=0}^{\infty} (-1)^n { i \choose n} C_{\text{S}_r^n} \right] \left( \frac{ \la \bf{21} \ra }{m} \right)^{2s-i}\left( \frac{x \la \bf{2}3 \ra \la 3 \bf{1} \ra}{m^2} \right)^{i}
\\ &\equiv (-1)^{2s} \frac{\k m x^2}{2 m^{2s}} \left[ g_0^r \la \bold{2} \bold{1} \ra^{2s} + g_1^r \la \bold{2} \bold{1} \ra^{2s-1} \frac{x \la \bold{2} q \ra \la q \bold{1} \ra}{m} + \cdots + g_{2s}^r \frac{(x \la \bold{2} q \ra \la q \bold{1} \ra)^{2s}}{m^{2s}} \right]
\eld
\el
Comparing the coefficients $g_i$ of eq.\eqc{eq:WCeff2gi} with the above formula, we can conclude that the Wilson coefficients $C_{\text{S}_r^n}$ used in eq.\eqc{eq:EFT3ptNew} is equivalent to $C_{\text{S}_{eff}^n}$; $C_{\text{S}_r^n} = C_{\text{S}_{eff}^n}$.

One advantage of the representation eq.\eqc{eq:EFT3ptNew} is that evaluation of cuts become simple as the sum over $2s$ intermediate states of a spin-$s$ particle can be substituted by a sum over 2 intermediate states of a spin-$\half$ particle. This property follows from the following identity.
\bl
\bld
\sum_{\bf{P}^{2s}} \CM_1(\bf{1},\bf{P},q_1,z_1)^{2s} \CM_2(\bf{P},\bf{2},q_2,z_2)^{2s} = \left[ \sum_{\bf{P}} \CM_1(\bf{1},\bf{P},q_1,z_1) \CM_2(\bf{P},\bf{2},q_2,z_2) \right]^{2s}
\eld
\el
In the above identity, $\CM_1$ and $\CM_2$ are arbitrary expressions bilinear in the massive spinor-helicity variables schematically written as \textbf{bold} variables. 
 This identity can be proved by writing the sum as the sum over overcomplete basis of spin coherent states. 

Another advantage of the expression eq.\eqc{eq:EFT3ptNew} is that it allows us to straightforwardly take the infinite spin-limit and connect to one-particle EFT three-point amplitude. 
Formally writing $\bar{u}u = 1$ and suppressing the subscript $s$ of $S^\m_s$,
\bl
\bld
\lim_{s \to \infty }M_s^{2 \eta} &= \lim_{s \to \infty} \frac{\k m x^{2\eta}}{2} \oint \frac{dz}{2\pi i z} \left( \sum_{n=0}^\infty C_{\text{S}_r^n} z^n \right) \left( 1 - \frac{1}{2s} \eta \frac{q \cdot S}{m z} \right)^{2s}
\\ &= \frac{\k m x^{2\eta}}{2} \sum_{n=0}^\infty \frac{C_{\text{S}_r^n} }{n!} \left( - \eta \frac{q \cdot S}{m} \right)^n \,,
\eld
\el
which is the one-particle EFT amplitude eq.\eqc{eq:1bd3ptAmp}.

\newpage

%


\bibliography{mybib}{}
\bibliographystyle{JHEP}

\end{document}